\documentclass[a4paper,fleqn,usenatbib]{mnras}

\usepackage{newtxtext,newtxmath}
\usepackage[T1]{fontenc}
\usepackage{ae,aecompl}

\usepackage{graphicx}	% Including figure files
\usepackage{amsmath}	% Advanced maths commands
\usepackage{amssymb}	% Extra maths symbols

\title[Aegaeon, Methone, Anthe, Pallene]{Long-term Evolution and 
Stability of Saturnian Small 
Satellites: Aegaeon, Methone, Anthe, and Pallene.}

\author[Mu\~noz-Guti\'errez \& Giuliatti Winter]{
M. A. Mun\~oz-Guti\'errez,$^{1}$\thanks{E-mail: mmunoz.astro@gmail.com (MAM)}
and S. Giuliatti Winter$^{1}$
\\
$^{1}$Universidade Estadual Paulista - UNESP, Grupo de Din\^{a}mica 
   Orbital e Planetologia, Av. Ariberto Pereira da Cunha, 333, 
   Guaratinguet\'a-SP, 12516-410, Brazil\\
}

\date{Accepted XXX. Received YYY; in original form ZZZ}

\pubyear{2016}

\begin{document}
\label{firstpage}
\pagerange{\pageref{firstpage}--\pageref{lastpage}}
\maketitle


\begin{abstract}

Aegaeon, Methone, Anthe, and Pallene are four Saturnian small 
moons, discovered by the Cassini spacecraft. Although
their orbital characterization has been carried on by a number 
of authors, their long-term evolution has 
not been studied in detail so far. In this work,
we numerically explore the long-term evolution, up to $10^5$ yr, 
of the small moons in a system formed by an oblate Saturn and the five largest
moons close to the region: Janus, Epimetheus, Mimas, Enceladus,
and Tethys. By using frequency analysis we determined the stability
of the small moons and characterize, through diffusion maps, the dynamical 
behavior of a wide region of geometric phase space, $a$ vs $e$, 
surrounding them. Those maps could shed light on the
possible initial number of small bodies close to Mimas, and help to 
better understand the dynamical origin of the small satellites.
We found that the four small moons are long-term stable and no mark of chaos
is found for them. Aegaeon, Methone, and Anthe 
could remain unaltered for at least $\sim0.5$Myr, given the current 
configuration of the system. They remain
well-trapped in the corotation eccentricity resonances 
with Mimas in which they currently librate. However, 
perturbations from nearby resonances, such as Lindblad 
eccentricity resonances with Mimas, seem responsible for largest 
variations observed for Methone and Anthe. 
Pallene remains in a
non-resonant orbit and it is the more stable, 
at least for 64 Myr. Nonetheless, it is affected by a quasi-resonance 
with Mimas, which induces long-term orbital oscillations of its 
eccentricity and inclination.

\end{abstract}


%%%%%%%%%%%%%%%%%%%%%%%%%%%%%%%%%%%%%%%%%%%%%%%%%%

%%%%%%%%%%%%%%%%% BODY OF PAPER %%%%%%%%%%%%%%%%%%

\section{Introduction}

The Cassini space mission has been a highly successful 
endeavor that helped to expand significantly our knowledge and 
understanding of Saturn and its environment, both physically
and dynamically. 
Among other important results,
several new moons around Saturn were discovered thanks to images taken
by the spacecraft. Four of such small moons, now called
Aegaeon, Methone, Anthe, and Pallene, lie between 
the position of the co-orbital duet formed by Janus and Epimetheus, 
at $\sim152\,000$ km from Saturn centre, and roughly the inner edge of the
E ring, where the moon Enceladus orbits, at $\sim238\,000$ km from Saturn 
centre. Those small bodies
remain a challenge to be successfully characterized due to the 
complex dynamical
environment they inhabit, nonetheless, such dynamical 
environment has started to be explored. In the present work, we review
our current knowledge of the four small moons, and explore their dynamical
environment and evolution on a wider and longer basis than current
studies did.\\

Aegaeon was discovered and reported by \citet{Hedman10}. 
It is a small moon of roughly 
0.33 km in diameter \citep{Thomas13} that lies in the 
middle of a prominent arc structure, located inside and
close to the inner edge of the
G ring. The G ring is a faint and thin ring formed 
by $\mu$m to cm sized particles \citep{Hedman07}, which extends from 
$\sim165\,000$ to $\sim175\,000$ km from Saturn centre \citep{Horanyi09}. 
Some particles forming the G ring presumably 
emanate from collisions of micrometeoroids with the
largest objects of the region; like Aegaeon itself, however, 
nowadays the origin
of the ring remains uncertain. As already mentioned, the G ring 
possesses a unique bright 
arc located close to 
its inner edge, at $\sim167\,500$ km from Saturn centre, which covers 
$60^{\rm o}$ in longitudinal extend, while has just $\sim$250 km 
of radial width. The arc is located at the 7:6 corotation eccentricity 
resonance (CER) with Mimas. It has been assumed that this resonance 
confines the longitudinal extent of the arc \citep{Hedman09}.\\  

Being the most prominent member of the arc of the G ring, 
Aegaeon orbits trapped at the 7:6 CER with Mimas, with a 
resonant argument given by 
$\varphi_{CER}=7\lambda_{Mimas}-6\lambda_{Aegaeon}-\bar{\omega}_{Mimas}$, 
where $\lambda$ and $\bar{\omega}$ are the mean longitude and 
longitude of pericenter, respectively. This resonant argument 
librates with amplitude 
$\sim10^\circ$ around $\sim180^\circ$.
However, Aegaeon's longitude deviates from the expected position,
if one only considers its $\sim10^\circ$ libration due to the CER. 
This small deviation has a 
period of $\sim$70 yr, similar to the period of the Mimas longitude libration
due to a 4:2 mean motion resonance (MMR) with Tethys \citep{Hedman10},
therefore Tethys perturbs the orbit of Aegaeon through its
interaction with Mimas.\\

Furthermore, variations in Aegaeon's eccentricity and inclination,
not related to the CER, indicate a possible influence of other resonances,
particularly the 7:6 Inner Lindblad Resonance (ILR), which resonant
argument is given by 
$\varphi_{ILR}=7\lambda_{Mimas}-6\lambda_{Aegaeon}-\bar{\omega}_{Aegaeon}$.
This resonant argument librates, although not so tightly 
as the argument of the CER,
with an amplitude of $\sim90^\circ$ around zero 
\citep{Hedman10}; therefore the ILR could be inducing second or 
higher order perturbations, not well
characterized so far.\\

Methone is a small moon of $\sim$1.45 km in diameter \citep{Thomas13}. 
First reported by \cite{Porco05}, it is located at 
$\sim194\,200$ km from Saturn centre. It inhabits inside an arc of 
dust for which most probably it is the source \citep{Sun17}. 
The arc's longitudinal 
extend is $\sim10^\circ$, roughly centered at the small moon. 
Both Methone and its arc orbit near the 15:14 CER and the 15:14 
outer Lindblad resonance (OLR) with Mimas \citep[both resonances are 
separated by just
4 km,][]{Elmoutamid14}. Both resonant arguments of the 15:14
CER and OLR librate with amplitude $\sim90^\circ$ around
zero. However,
the arc confinement is consistent with the maximum extension 
permitted by the 15:14 CER with Mimas alone \citep{Hedman09}.\\

Anthe is a tiny moon of an estimated $\sim$0.5 km in diameter; 
a better define measurement has not been carried out to date,
due to the lack of resolution in Cassini images 
\citep{Thomas13}. Anthe is located at
$\sim197\,600$ km from Saturn centre. It was first reported by
\cite{Cooper08}, whom characterized its short-term dynamics.
They found that Anthe orbits close to an 11:10 outer resonance with Mimas, 
with two librating arguments, one of which is associated with
the 11:10 CER with Mimas,
$\varphi_{CER}=11\lambda_{Anthe}-10\lambda_{Mimas}-\bar{\omega}_{Mimas}$,
while the second one is given by
$\varphi_2=11\lambda_{Anthe}-10\lambda_{Mimas}-\bar{\omega}_{Anthe}
-\Omega_{Anthe}+\Omega_{Mimas}$. Both periods of libration for these
arguments are 761 days \citep{Cooper08} and librate with amplitude
$\sim90^\circ$ around zero.
Likewise Aegaeon and Methone, Anthe lies at the centre 
of an arc of dusty material that extends
for $\sim20^\circ$, consistent with the confinement produced by the
11:10 CER with Mimas \citep{Hedman09}.\\

Both arcs of Methone and Anthe have been deeply studied recently by
\cite{Sun17}. They explored the evolution of 
$\mu$m sized particles
which origin is the ejecta produced by collisions of interplanetary 
dust particles (IDP) on the surfaces of
the small moons. They then explored the dynamical evolution of the ejected 
particles
considering an oblate Saturn, the gravitational perturbations of Mimas,
Enceladus, and Methone or Anthe, depending on the system, plus a number
of non-gravitational effects, such as solar radiation pressure, Lorentz
force, and plasma drag. The authors are able to reproduce some of the
properties of both arcs, such as their longitudinal extend and optical 
depth,
although dust density in the simulations is an order of magnitude 
smaller compared to estimations derived from observations. 
According to \cite{Sun17},
particles smaller than 5 $\mu$m are quickly removed from the arcs due 
mainly to the action of plasma drag,
while larger particles could survive up to 100 yr in the arcs
before recolliding with the parent small moon or leaving the arc.\\

Regarding Pallene, also first reported by \cite{Porco05}, it 
is a small moon of 
diameter $\sim$2.23 km \citep{Thomas13}, orbiting at 
$\sim212\,300$ km from Saturn centre. It shares its orbit with a 
faint full ring of dusty material. This ring is only
visible at extremely high phase angles, implying that it is
formed by $\mu$m sized particles. The Pallene ring is very thin, 
with a vertical width of just $\sim$50 km. For comparison, Pallene 
achieves vertical excursions  
of $1\,300$ km above Saturn's equatorial plane,
due to its relatively large inclination. The ring is therefore
tightly constrained to the orbital plane of Pallene, sharing
the same inclination of the small moon \citep{Hedman09}.
Regarding their dynamics, \cite{Spitale06} suggested that both the 
moon and its ring were probably in a 19:16 inner mean 
longitude resonance with Enceladus, with resonant argument given by
$\theta=19\lambda_{Enc}-16\lambda_{Pall}-\bar{\omega}_{Pall}-
2\Omega_{Pall}$. Meanwhile, \cite{Callegari10} argued for
a secular resonance relating Pallene and Mimas, with resonant
argument given by $\theta=\bar{\omega}_{Pal}-\bar{\omega}_{Mim}-
\Omega_{Mim}+\Omega_{Pal}$. Nonetheless, both arguments circulate,
with different periods, suggesting that Pallene is most probably
evolving out of any strong resonance. We provide some details  
in sections \ref{sec:shortt} and \ref{sec:longt}.\\

The short-term dynamics of the four small moons have been explored 
in a number
of works, mainly those concerning their discovery and orbit determination
from Cassini raw images. As we have recalled, Aegaeon, Methone 
and Anthe orbit near CER with Mimas, while Methone and Anthe
also lie close to Lindblad resonances with Mimas. CER and Lindblad 
resonances are separated by just some km and are very close to the mean
semimajor axis of Methone and Anthe. Motivated by this,
\cite{Elmoutamid14} proposed a simplified analytical model, named 
CoraLin model, 
to explore the dynamics of test particles close to both
CER and Lindblad eccentricity resonances (LER), which can be applied, 
under some assumptions, to these small moons.
The CoraLin model is a simplification of the Hamiltonian
for a restricted three body problem, where the large perturbing moon
is eccentric and the test particle orbits near a MMR of the 
form $m+1:m$, where $m$ is an integer. The model depends upon just two 
free parameters. The first one, $D_C$, is related to the
separation distance between the positions of the CER and LER, while 
the second parameter, $\epsilon_L$, is related to the mass and eccentricity 
of the perturbing satellite. $D_C=0$ means that both CER and LER lie 
at the same position,
and the Hamiltonian is integrable in this case. For $D_C>2$ both
resonances are separated enough as to not interfere with each other and
the Hamiltonian is quasi integrable. 
Intermediate cases, $0<D_C<2$ leads to strong coupling between
the resonances, which in turn leads to a chaotic evolution of
the orbits.\\

The results from applying the CoraLin model to the small moons,
by inspecting the corresponding Poincar\'e surfaces of section, suggest
that Anthe is closer to the chaotic region than any of the other small 
moons. Methone is farther from the chaotic region than Anthe,
while Aegaeon is the more stable, for being
far enough from the chaotic region. 
Those small moons form a distinctively family of objects
which dynamics is dominated by CER and LER coupling resonances
\citep{Cooper15}.\\

The only long-term exploration so far regarding the dynamics of the 
small moons (except for Aegaeon), was carried out by \cite{Callegari10}.
They performed $\sim60\,000$ yr numerical integrations of the orbits
of Methone, Anthe and Pallene under the influence of Saturn's oblateness
until $J_4$, and the gravitational perturbations from Mimas, 
Enceladus, Tethys, 
Dione, Rhea, and Titan. They found a high sensitivity on initial 
conditions for both Methone and Anthe, suggesting that their evolution
could be chaotic, while Pallene seems to be very stable. The authors
suggest a possible resonance involving the longitudes of pericenter 
and ascending nodes of Pallene and Mimas, however, as previously
noted, such resonant argument does not librate, but
circulates with a long period, as we will verify in sections 
\ref{sec:shortt} and \ref{sec:longt}.\\

Regarding their physical characteristics, the Cassini images have revealed
smooth surfaces for two moons, Methone
and Pallene, for which there are enough resolution in the images. 
It is expected that both Aegaeon and Anthe share 
the same feature, i.e. a smooth surface; this because of their 
proximity, size and
likely same formation process. Moreover, since the discovery of
the arc/ring structures associated with the small moons, the source
of the micrometric material forming such structures
has been assumed to be the escaped debris particles produced
when IDP collided on the surface of the small moons. Due to their 
small size, their escape velocity is very low, letting
almost all impact ejecta freely evolve out of the surface of
the small moon into
the arcs/ring. However, a considerable proportion of those dusty
particles end up colliding again with the moons, smoothly 
depositing themselves on the surface of the small moons, thus being likely 
responsibly for the even surfaces observed. Numerical 
evidence for such scenario can be found in \cite{Sun17}.\\

In this paper we explore the long-term evolution of the four small
moons, exceeding any previous integration of their orbits. We also 
explore, with thousands of test particles, a wide region of the 
geometric phase space of semimajor axis, $a$, \textit{vs}  
eccentricity, $e$, surrounding the small moons, to gain a global
understanding of the dynamical behavior and evolution of such
regions where arcs/ring structures coexist.\\

This paper is organized as follows: in section \ref{sec:methods} we
describe the frequency analysis technique and the simulations 
used in this work. Section \ref{sec:shortt} 
is devoted to the short-term evolution of the small moons, while
in section \ref{sec:global} we present 
several diffusion maps, based on frequency analysis, which characterize 
the dynamical behavior of all the regions of interest. In section
\ref{sec:longt} we present our main results regarding the
long-term evolution of the small moons. Finally, in section \ref{sec:conc}
we enumerate our main conclusions.\\

\section{Methods and Numerical Simulations}
\label{sec:methods}

\begin{table*}
\centering
\caption{Summary of physical parameters of the moons.}
\label{tab:tabpar}
\begin{tabular}{@{}lcccc}
\hline
\hline
Name & GM (km$^3$ s$^{-2}$) & $\rho$ (g cm$^{-3}$) & $R_m$ (km) & 
Reference \\
\hline
Janus & 1.2656324971531704E-01 & 0.63 & 89.2 & \cite{Thomas13} \\
Epimetheus & 3.512421991952764E-02 & 0.64 & 58.2 & \cite{Thomas13} \\
Aegaeon & 1.2606299971854121E-09 & 0.54 & 0.33 & \cite{Thomas13} \\
Mimas & 2.502784093954375E+00 & 1.15 & 198.2 & \cite{Cooper15} \\
Methone  & 6.261297469640338E-08 & 0.31 & 1.45 & \cite{Thomas13} \\
Anthe & 2.334499994563023E-08 & 0.35 & 0.5 & \cite{Thomas13} \\
Pallene & 1.7980185560068351E-07 & 0.25 & 2.23 & \cite{Thomas13} \\
Enceladus & 7.211597878640501E+00 & 1.6 & 252.6 & \cite{Cooper15} \\
Tethys & 4.121706150116760E+01 & 0.956 & 537.5 & \cite{Cooper15} \\
\hline
\end{tabular}

\medskip
\end{table*}
\begin{table}
\centering
\caption{Saturn parameters.}
\label{tab:tabSat}
\begin{tabular}{@{}lcc}
\hline
\hline
Constant & Value & Reference \\
\hline
GM (km$^3$ s$^{-2}$) & 3.793120706585872E+07 & \cite{Cooper15}\\
$R_S$ (km) & 60,330 & \cite{Cooper15} \\
$J_2$ & 1629.054382E-05 & \cite{Hedman10} \\
$J_4$ & -93.6700366E-05 & \cite{Hedman10} \\
$J_6$ & 8.6623065E-05 & \cite{Hedman10} \\
\hline
\end{tabular}
\medskip
\end{table}

\subsection{Frequency Analysis}

Frequency analysis is a powerful technique to quantify the weak chaotic
behavior of a dynamical system with arbitrary degrees of freedom. 
Since its original formulation by \cite{Laskar90, Laskar93}, who develop it 
in order to prove the chaotic nature of the solar system's secular 
evolution, it has been successfully applied to a wide variety of 
dynamical problems, going from planetary sciences 
\citep{Nesvorny98,Robutel01}
to galactic dynamics \citep{Papaphilippou98,Valluri98},
and even through fundamental particle physics 
\citep{Nadolski03,Papaphilippou14}.\\

The frequency analysis algorithm developed by Laskar, look for 
the fundamental frequencies
of the system resulting from a numerical integration. 
A brief sketch of the method and, at
the same time, a justification for its use,
can be stated as follows \citep{Laskar92}:
consider the Hamiltonian 
of an integrable
dynamical system, $H(J,\theta)$; once the system 
is reduced to action angle variables, $(J_j,\theta_j)$, the 
Hamiltonian will depend only on the actions, $H(J,\theta)=H_0(J_j)$, where 
$j=1,2,...,n$ for a system of $n$ degrees of freedom. 
We know that actions are constants 
of the motion, $\dot J_j=0$, while angles evolve according to:
\begin{equation}
\dot{\theta_j}=\frac{\partial H_0(J_j)}{\partial J_j}=\nu_j(J),
\label{eq:tep}
\end{equation}
where the $\nu_j(J)$ are the fundamental frequencies of the motion.\\

In numerical experiments, we will rarely work on action angle 
variables, nonetheless, we can still consider some close related 
dynamical variables, say $z'_j$, of the form: 
$z'_j=J'_j\,\exp(i\theta'_j)=f(z_1,z_2,...,z_n)$, where 
$z_j=J_j\,\exp(i\theta_j)$, this is, 
any variable will be a certain function, $f$, close to unity, 
that depends on the actual actions and angles of the system. Although 
$(J'_j,\theta'_j)$ are not the action angle variables, the analysis of
$z'_j(t)$ will still give us the fundamental frequencies, $\nu_j$, since 
for periodic and quasi periodic motions, these frequencies 
will remain constant. \\

If we recover the fundamental frequencies in different
intervals of a numerical integration, we will be able to determine the
existence of periodic motion, if frequencies remain almost constant,
or on the contrary, if largely variable frequencies are found,
the implication is 
an irregular, maybe chaotic motion. This is because, by inverting
Eq. \ref{eq:tep}, we can state that $J=F(\nu_1,\nu_2,...,\nu_n)$, 
therefore
variable frequencies imply variable actions.\\

We are not interested in recovering the analytical representation of
particles orbits in our region of interest. On the
contrary, we are interested in the variations of their motions due
to different perturbations, thus we focus on the global
variations of the frequencies and not in their specific values.   
In this work we calculate the main frequency, $\nu$, of the
quantity $z'_j=a(t)\exp(i\lambda(t))$, from a time
series dataset resulting from a numerical integration. To obtain the 
frequencies we use
the frequency modified Fourier transform (FMFT) algorithm, described in
\cite{Sidli96}, made publicly available by the author
\footnote{https://www.boulder.swri.edu/\textasciitilde davidn/fmft/fmft.html}.\\

According to \cite{Robutel01}, the main
frequency of $z'(t)$ will be related to
the mean motion, $n$, of the orbit. A full representation of
$z'(t)$ through frequency analysis would be:
\begin{equation}
z'(t)=\alpha_0\exp(i\nu_0t)+\sum^N_j\alpha_j\exp(i\nu_jt),
\end{equation}
where $\nu_0=n$ in the case of pure Keplerian motion. As this is not
the case, both $|\alpha_0|$ and $\nu_0$ are close, but not equal, to
the initial semimajor axis and mean motion, 
respectively. Nonetheless, the amplitudes of the following terms are
100 to $1\,000$ times smaller for quasi-periodic trajectories, therefore, 
variations in the main frequency, $\nu_0$, will be enough to provide 
an estimation of the orbit's stability. 

\subsection{Numerical simulations}

In order to obtain a global dynamical perspective
of the region inhabited by Aegaeon, Methone, Anthe, 
and Pallene, we performed
short-term ($\sim$18 yr) numerical simulations 
of thousands of test particles that initially
cover the entire $a$ vs $e$ geometric phase-space, 
from the orbit of 
Janus-Epimetheus, to beyond the orbit of Enceladus. The choice 
of our integration time relies on a compromise, between the large 
volume of data produced and the requirement of enough orbital periods
to accurately calculate the main frequencies for our analysis. In
$\sim$18 yr Mimas performs $\sim6\,950$ orbital revolutions around
Saturn, while
Enceladus, close to the exterior limit of our maps, performs $4\,780$ 
revolutions, i.e. we obtain enough orbital revolutions for all particles 
for an accurate recovery of their frequencies.\\

Test particles are subject to gravitational perturbations from the oblateness 
of Saturn up to $J_6$ in zonal harmonics,
plus the five largest moons of the region, namely 
Janus, Epimetheus, Mimas, Enceladus, and Tethys, in a first set. 
In later simulations we consider the gravitational perturbations 
from Saturn and the large five moons, as well as the gravitational
perturbations from the four small moons. After the global exploration,
we zoomed in towards those regions closer to each small moon, performing 
simulations that include thousands of test particles
covering a smaller patch of the geometric $a$ vs $e$ 
phase-space, but in finer steps. We
provide the details for each simulation along with the results in the next 
section.\\

In section \ref{sec:longt}, the results for a single long-term 
($1\times10^5$ yr) numerical integration are reported,
where only the interaction between the five large moons and the four
small moons, orbiting an oblate Saturn, is considered.\\
 
All of the numerical simulations presented in this work were performed
using the B\"ulirsch-St\"oer integrator from the MERCURY package 
\citep{Chambers99}, where a toleration accuracy parameter was set to
$10^{-12}$ and an initial time-step of $0.1$ day was used.
Most of the simulations were performed at the Saturn Cluster 
belonging to the Group of Planetology and 
Orbital Dynamics of the Mathematics Department of the S\~ao Paulo State
University (UNESP).\\

The physical parameters of all the nine different moons used 
throughout this work are shown in Table \ref{tab:tabpar}, where
GM is the gravitational parameter, $\rho$ is the bulk density
of the body, and $R_m$ the mean radius. In Table \ref{tab:tabSat}
we show the parameters of Saturn.\\

\section{Short-term Dynamics}
\label{sec:shortt}

\begin{table*}
\centering
\caption{Initial conditions in geometric elements for the large moons.}
\label{tab:Tabinilg}
\begin{tabular}{@{}lccccc}
\hline
\hline
Parameter & Janus & Epimetheus & Mimas & Enceladus & Tethys \\
\hline
$a$ (D$_{Mim}$) & 0.816224821915373 & 0.816481357153813 & 1.00001997053844 & 1.282926463121003 & 1.588211135154469 \\
$a$ (km) & $151\,441.5372$ & $151\,489.1345$ & $185\,542.7053$ & $238\,032.8930$ & $294\,675.1058$ \\
$e$ & 0.006783798913451 & 0.009648290023497 & 0.01935393988488 & 0.004717750162711 & 0.000318152308649 \\
$i$ ($^\circ$) & 0.163561569028246 & 0.351528831169069 & 1.56783494576081 & 0.004011975196723 & 1.091284647139937 \\
$\bar\omega$ ($^\circ$) & 6.132967855641644 & 116.348018800562642 & 88.06783652593947 & 57.40490801677754 & 29.38263145675355 \\
$\Omega$ ($^\circ$) & 6.369960248480534 & 44.525088430924164 & 239.785329147143301 & 337.68709939428453 & 141.711912205570002 \\
$\lambda$ ($^\circ$) & 129.813146812610796 & 263.577236051360842 & 318.616779106079946 & 217.90585212208282 & 271.834917459084238 \\
\hline
\hline
\end{tabular}
\medskip
\end{table*}

\begin{table*}
\centering
\caption{Initial conditions in geometric elements for the small moons.}
\label{tab:Tabinism}
\begin{tabular}{@{}lcccc}
\hline
\hline
Parameter & Aegaeon & Methone & Anthe & Pallene \\
\hline
$a$ (D$_{Mim}$) & 0.90274553988027 & 1.046749724679189 & 1.065312819759956 & 1.144147730931239 \\
$a$ (km) & $167\,494.5047$ & $194\,212.8971$ & $197\,657.0752$ & $212\,284.0258$ \\
$e$ & 0.000312629006216 & 0.000781255131993 & 0.000980088315003 & 0.003986601194199 \\
$i$ ($^\circ$) & 0.002368099675087 & 0.010191700663056 & 0.018728031233752 &  0.181964040454351 \\
$\bar\omega$ ($^\circ$) & 240.784600532626655 & 101.055122930558355 & 196.379681976898155 & 320.254677876382459 \\
$\Omega$ ($^\circ$) & 246.490855111672857 & 150.999808094857514 & 115.172628200709838 & 131.420435093503784 \\
$\lambda$ ($^\circ$) & 325.21506542212154 & 205.403404370634803 & 159.569919065624163 & 293.49347042844056 \\
\hline
\hline
\end{tabular}
\medskip
\end{table*}

\begin{figure}
\includegraphics[width=\columnwidth]{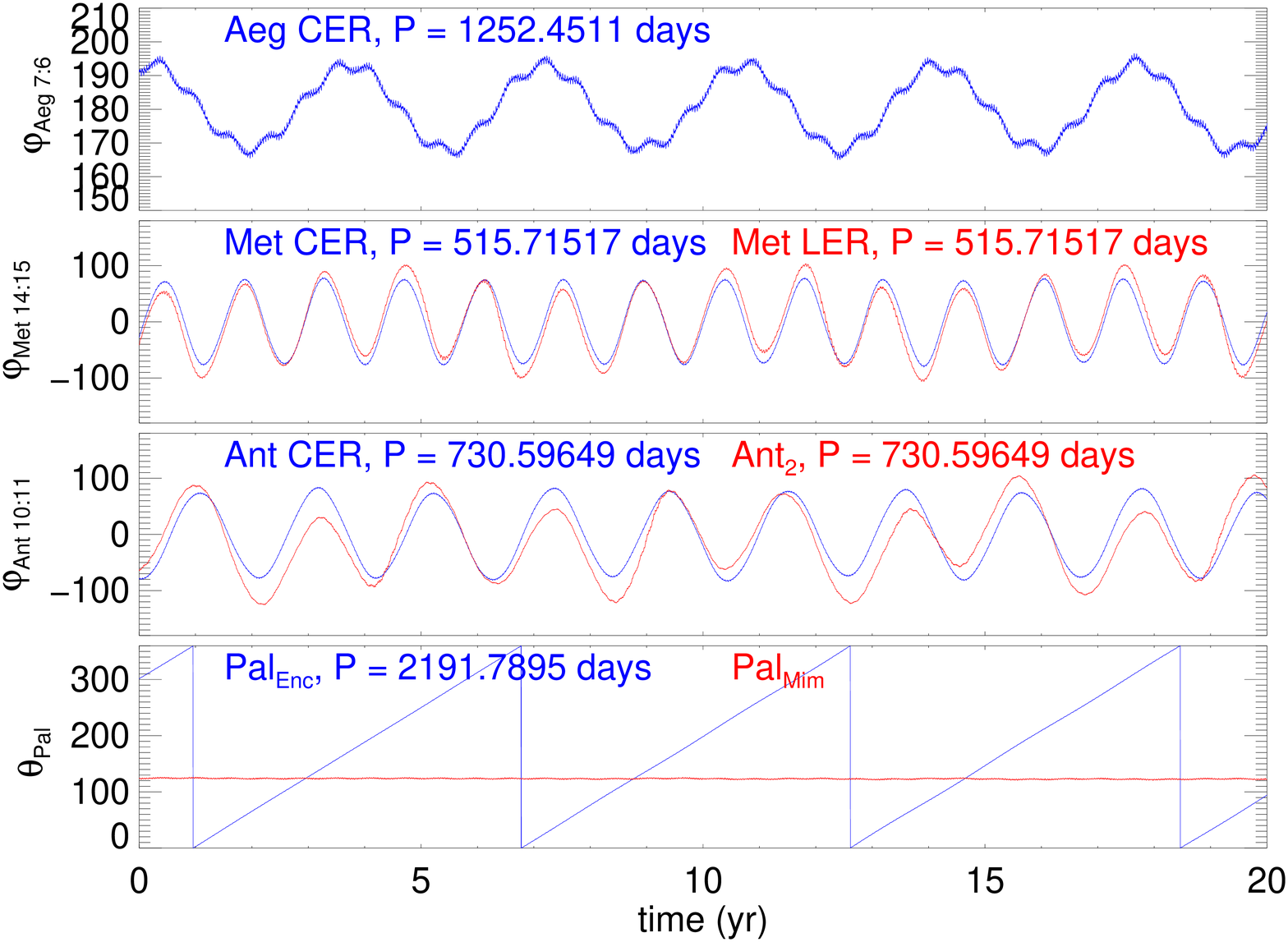}
\caption{Short-term evolution of different resonant arguments
of the small moons. Each panel shows the evolution for 20
yr of the arguments associated to the resonances in which 
the small moons are known or suggested to be trapped. Top panel 
for Aegaeon, second for Methone, third for Anthe, and the bottom 
panel for Pallene. See text for details.}
\label{fig:reso}
\end{figure}

To test the robustness of the initial conditions used in all of our 
simulations, we performed a detailed short-term integration, lasting
for 24 yr, in order to reproduce the resonant dynamics of
Aegaeon, Methone, and Anthe, as has been found
by previous authors. This simulation includes the nine moons 
presented in Table
\ref{tab:tabpar}, plus Saturn and its gravity coefficients given in 
Table \ref{tab:tabSat}. 
The initial conditions for all the bodies
were taken from the Horizons website for the Julian Day 
$2\,457\,601.5$, which corresponds to the 
date August 01, 2016. The geometric elements used as 
initial conditions for the nine objects which constitute the n-body 
part of the simulations, 
are shown in Tables \ref{tab:Tabinilg} (for the five large moons) 
and \ref{tab:Tabinism} (for the four small moons), where 
all variables have their usual meaning of: geometric semimajor axis, $a$, 
geometric eccentricity, $e$, geometric inclination, $i$, 
longitude of pericenter, 
$\bar\omega$, longitude of the ascending node, $\Omega$, and mean
longitude, $\lambda$. In Tables
\ref{tab:Tabinilg} and \ref{tab:Tabinism}, 
D$_{Mim}$ stands for the normalized geometric semimajor axis of Mimas, 
D$_{Mim}=1$, used as a reference
distance unit throughout this work. In physical units 
D$_{Mim}=1.85539\times10^5$ km.\\
 
The most important constraint for the dynamics
of the small moons is the resonant dynamics associated with several
resonances, among them the CER and LER 
with Mimas. In Fig. \ref{fig:reso} we show the short-term 
evolution of
the resonant arguments for Aegaeon, Methone, and Anthe resonances.
In addition, we plotted the evolution of the two arguments associated with
the quasi-resonances of Pallene with Enceladus and Mimas, as suggested 
by \cite{Spitale06} and \cite{Callegari10}, respectively.\\ 

\cite{Hedman10} found that Aegaeon is trapped in the 7:6 CER with Mimas,
with a libration argument $\phi_{Aeg}
=7\lambda_{Mim}-6\lambda_{Aeg}-\bar{\omega}_{Mim}$, a
period of libration around 180$^\circ$ of $\sim1\,264$ days, and 
amplitude of $\sim10^\circ$. We found (top panel of Fig. \ref{fig:reso}) 
a libration period of
$\sim1\,252.5$ days, a difference of only 0.9\% with the previous
result, and the same libration
amplitude of 10$^\circ$ around 180$^\circ$.\\

For Methone, both \cite{Spitale06} and \cite{Jacobson06} found a
libration argument $\phi_{Met}^{LER}
=15\lambda_{Met}-14\lambda_{Mim}-\bar{\omega}_{Met}$, a 
period of libration of $\sim$450 days, and an amplitude of the
residual longitude of
$\sim5^\circ$. This argument corresponds to the 15:14 LER with Mimas.
Nonetheless, \cite{Hedman09} also found that the argument
associated to the 15:14 CER with Mimas, $\phi_{Met}^{CER}
=15\lambda_{Met}-14\lambda_{Mim}-\bar{\omega}_{Mim}$, also
librates with the same period and amplitude as the one associated
with the LER. This means than Methone is perturbed by both the CER and
the LER with Mimas, which are separated by just 4 km. We 
found that effectively both arguments for the CER (blue curve in second panel
of Fig. \ref{fig:reso}) and the LER (red curve in same panel) of Methone, 
librate, and we have normalized the
two arguments for them to librate around zero, in order to compare their
amplitudes, which is around 90$^\circ$. Their period of libration is
of $\sim$515.7 days,
a difference of $\sim$14\% with previous results. Such 
differences could
be the result of the updated orbital data and masses we have 
used. An 
evidence in this sense is the recent result found by \citet{Sun17}
for the Methone libration period of the CER. They found this period
to be $\sim520$ days and an amplitude of the residual longitude of
$\sim5^\circ$. The result of \cite{Sun17} is in better agreement 
with ours, since they used updated data from the Horizons system
of the Jet Propulsion Laboratory, as we did.\\

In the case of Anthe, \cite{Cooper08} found two resonant arguments that 
librate, one is the associated with the 11:10 CER with Mimas,
$\phi_{Ant}
=11\lambda_{Ant}-10\lambda_{Mim}-\bar{\omega}_{Mim}$, and the second
argument is given by $\phi_{Ant_2}=11\lambda_{Ant}-10\lambda_{Mim}
-\bar{\omega}_{Ant}
-\Omega_{Ant}+\Omega_{Mim}$. For both arguments, \cite{Cooper08}
found a period of libration of $\sim$761 days and an amplitude of
78$^\circ$. We find for both arguments (blue and red curves in
third panel of Fig. \ref{fig:reso}) a period of $\sim731$ days, 
a difference of
just 4\% with the previous result, with an amplitude of libration of 
$\sim90^\circ$.
\cite{Hedman09} and \cite{Sun17} argue for the 11:10 CER of Anthe with
Mimas to be responsible for the confinement of the Anthe's arc,
being then the most relevant for the evolution of the moon and
its environment.\\

Finally, although Pallene is not in resonance neither with Mimas or
Enceladus, both \cite{Spitale06} and \cite{Callegari10} suggest
two different arguments that circulate and could indicate a 
quasi-resonance of Pallene with any of those large moons. 
First, \cite{Spitale06} suggest the
argument: $\theta_{Pall}^{Enc}=19\lambda_{Enc}-16\lambda_{Pall}-
\bar{\omega}_{Pall}-2\Omega_{Pall}$, a 19:16 inner mean
longitude resonance between
Pallene and Enceladus, as responsible for the long-term 
variations in the orbital elements of Pallene. On the other hand,
\cite{Callegari10} argue for an argument given by 
$\theta_{Pall}^{Mim}=\bar{\omega}_{Pal}-\bar{\omega}_{Mim}-
\Omega_{Mim}+\Omega_{Pal}$, which implies a relation between 
Pallene and Mimas, based on the high stability found by the authors
in their long-term simulations. They obtain a period of circulation
of $\sim4\,400$ yr. We find that in fact both cited
arguments circulate, where the one involving Pallene and Enceladus
has a period of circulation of $\sim2\,192$ days (blue curve in bottom panel
of Fig. \ref{fig:reso}), while the second argument, involving Mimas and
Pallene, although at first glance could seem as librating (see red curve
in Pallene's panel), it actually
circulates but with a much larger period, not visible in this
short-time scale plot. We show the circulation of this second argument in
section \ref{sec:longt}, for which we find a period of $\sim4\,762.21$ yr, 
a difference of $\sim$7.6\% compared with \cite{Callegari10}
work.\\

\section{A Global Dynamical Perspective Through Frequency Analysis
Maps}
\label{sec:global}

\subsection{The Full Region Around Mimas}

\begin{figure*}
\includegraphics[width=17cm]{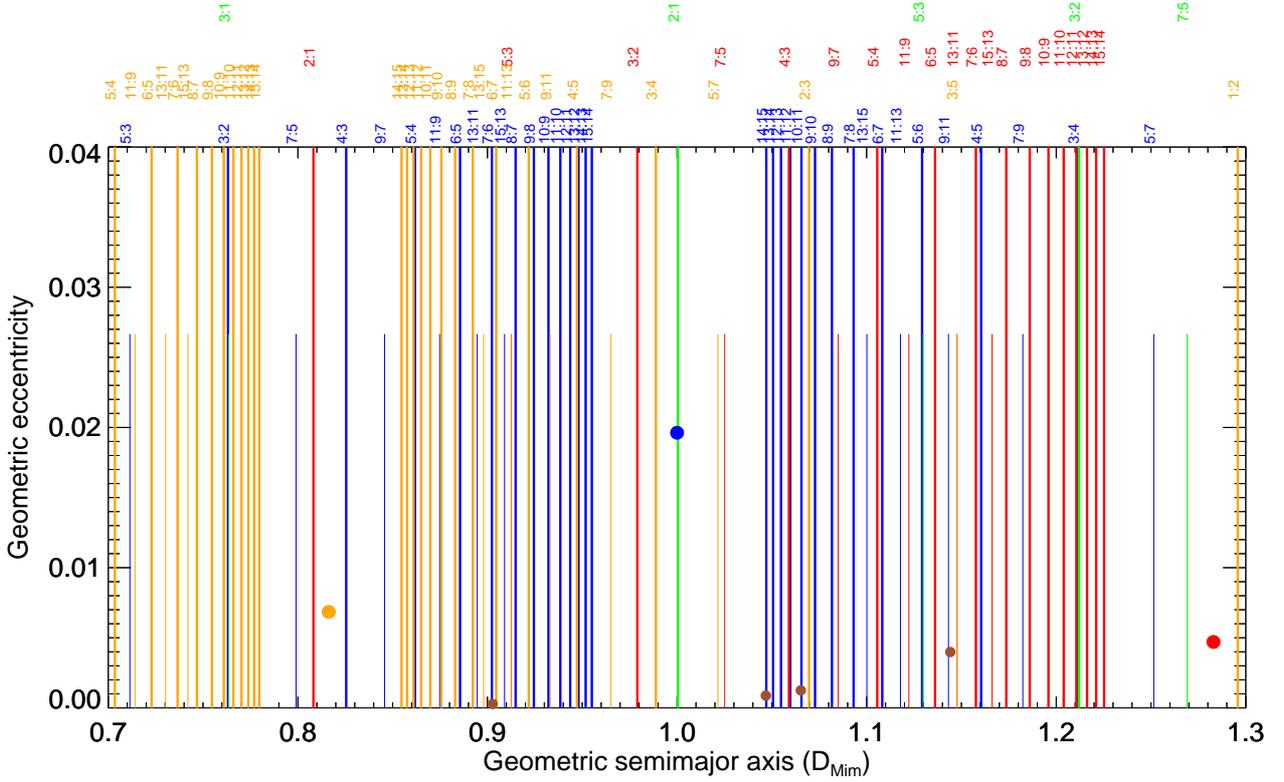}
\caption{Location of first and second order mean motion resonances
in the region. We plot colored vertical lines at the position of
interior and exterior first order MMRs (thick lines from bottom to top)
and second order MMRs (thin lines from bottom to middle), up to 
the ratio $p/q=15/14$,
inside the region of phase-space, of geometric 
semimajor axis \textit{vs} geometric eccentricity, explored in this
work. Color code is as follows: orange lines stand for MMRs with Janus
(moon location also shown as filled orange circle), blue lines for 
MMRs with Mimas
(moon as blue circle), red lines for MMRs with Enceladus (moon as
red cirlce), and green lines for MMRs with Tethys (not shown in the plot). The 
corresponding MMR ratios are shown in an adequate color at the top of
the figure. Small filled brown circles indicate the location of
the four small moons, from left to right: Aegaeon, Methone, Anthe, and
Pallene.}
\label{fig:AllRes}
\end{figure*}

\begin{figure*}
\includegraphics[width=17cm]{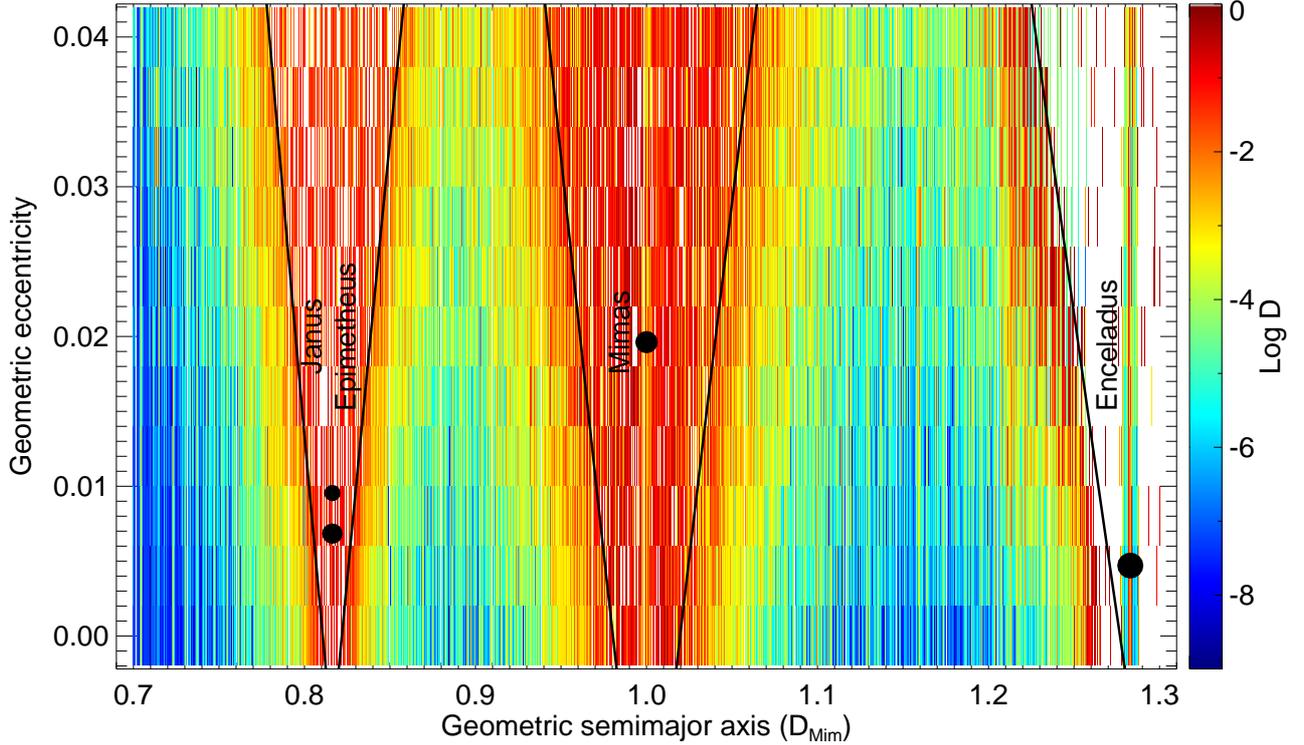}
\caption{Diffusion map for the full region of interest. For this map,
along with test particles, only the five large moons were included 
in the simulation, although Tethys is not shown due to its farther 
location. The diffusion parameter, $D$, is obtained from Eq. \ref{Eq:diffp}
after a frequency analysis is performed for each test particle. 
A rectangle, colored according to the value of Log($D$), 
is plotted centered at each initial condition in the 
phase-space plane of geometric semimajor axis \textit{vs} geometric 
eccentricity. Bluest
regions correspond to the more stable orbits, while redder zones 
correspond to unstable trajectories. Some particles do not survive 
the whole simulation and for them we plotted a white rectangle, same 
color as the background.}
\label{fig:FReg5bb}
\end{figure*}

\begin{figure*}
\includegraphics[width=17cm]{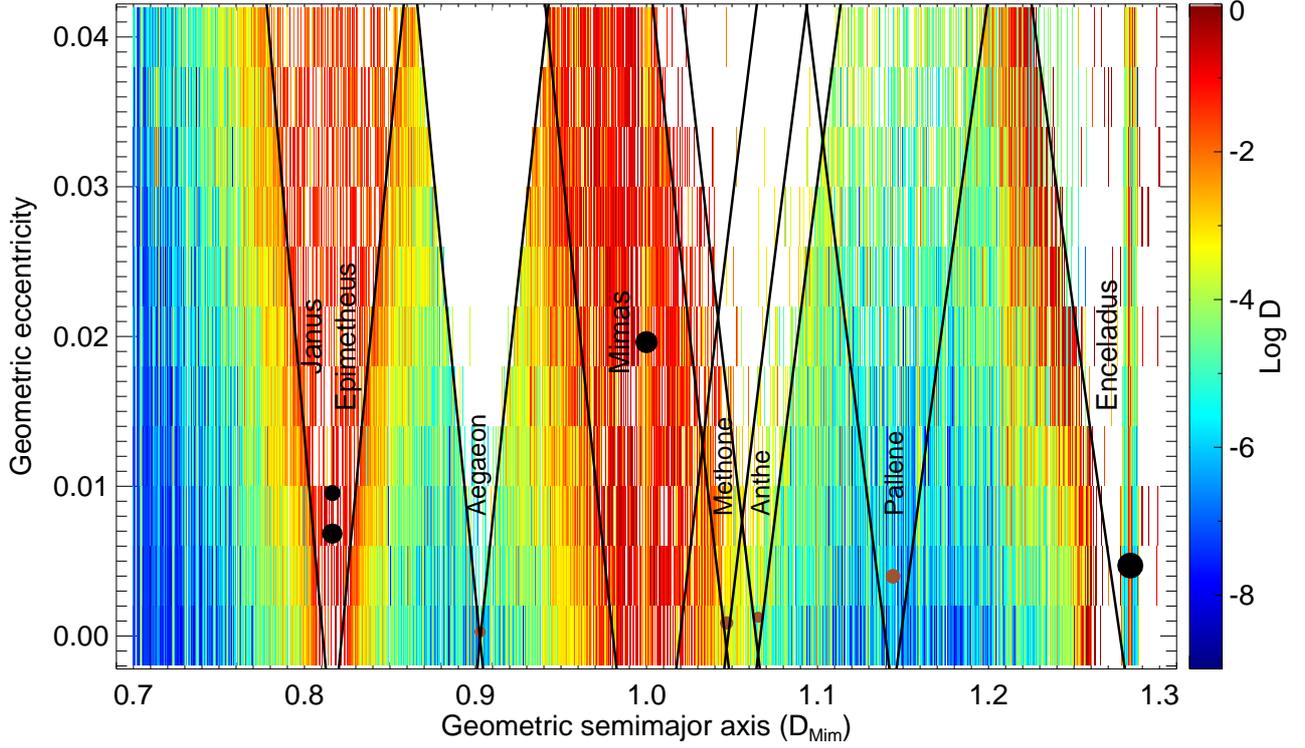}
\caption{Diffusion map of the region of interest including all nine
moons as n-bodies. As in Fig. \ref{fig:FReg5bb}, scale colored 
rectangles are plotted for each initial condition according to 
the $D$ value of the particle. Wider white regions, were particles 
are removed before the end of the simulation, are now present due to
collisions with the small moons.}
\label{fig:FRegFull}
\end{figure*}

\begin{figure*}
\includegraphics[width=17cm]{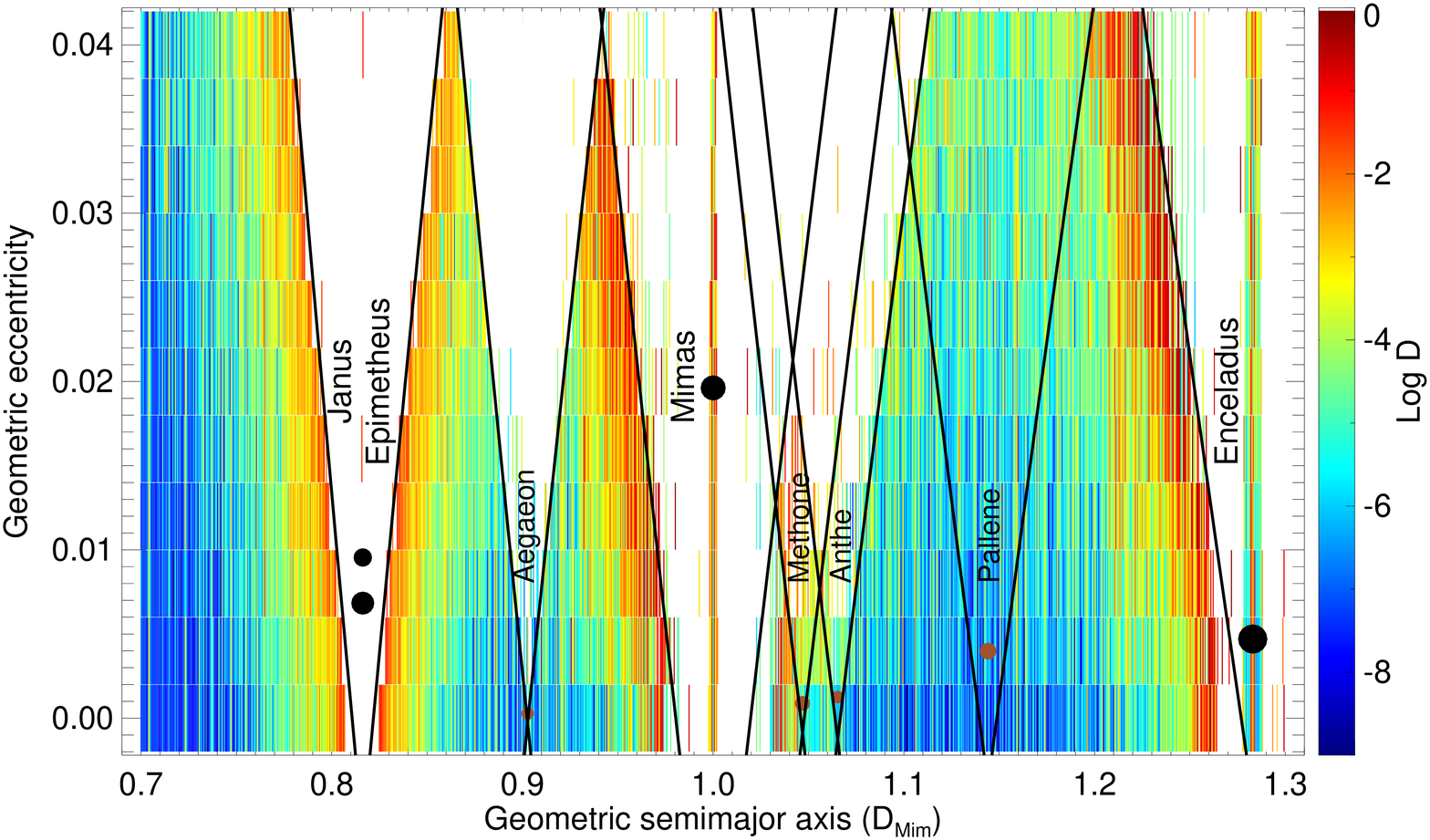}
\caption{Same as Fig. \ref{fig:FRegFull} but with the initial
inclination of all moons set to zero.}
\label{fig:FRegZero}
\end{figure*}

The aim of this section is to delineate
the lookalike of a wide dynamical region around Mimas, going from
0.7 to 1.3 $D_{Mim}$, to better understand the influences that
can drive the evolution of the small moons and their ring/arcs. In
Fig. \ref{fig:AllRes} we plot the location of all first and second
order MMRs with Janus, Mimas, Enceladus, and Tethys, present in this
region of the Saturnian system. All this resonances are somewhat
responsible for the shaping of structures, cleaning or maintaining 
of the particles in the region, which are subject to gravitational 
forces only. Although it is well-known that all Aegaeon, Methone,
Anthe, and their associated arcs lie in CER with Mimas, here we 
provided a global framework for further detailed explorations 
near any specific MMR of interest, where other kinds of dynamics could 
better fit to the particular case, such as corotation or Lindblad 
resonances.\\

The global dynamical picture of the full region was obtained 
from the detailed numerical 
integration during $\sim$18 yr of $11\,000$ test
particles distributed in a rectangular grid, covering the geometric 
phase-space of $a$ \textit{vs} $e$. In $a$, 
initial conditions are taken from 
0.7 to 1.3 D$_{Mim}$ in steps of $6.0\times10^{-4}$.
In $e$ we cover from 0 to 0.04 in steps of $4\times10^{-3}$.
In all this work we explore the evolution of particles initially
in the Saturn's
equatorial plane (thus all test particles
have initially zero inclination). As for the remaining 
orbital elements of each particle,
$\bar{\omega}$, $\Omega$, and $\lambda$, they are set to zero,  
following the approach of \cite{Robutel01} in their global exploration
of the dynamics of the solar system. Indeed, the variations of the angular 
elements ($\bar{\omega}$, $\Omega$, and $\lambda$) do not change significantly
the results for the diffusion of the particles outside resonances (which
constitutes most of the dynamical space in all our experiments). 
Such variations lead only to different capture rates for
particles which are initially close to MMRs and therefore
are subject to get trapped. In this sense, angles' values are important
mainly because they determine the width of the libration regions 
of MMRs. Nonetheless, in our case, as we only explored very low 
eccentricities, the width of MMRs is, as we will see, very thin,
thus the contribution from variations of the angles is negligible.\\

All test particles are subject to the non-spherical potential of 
Saturn up to an order $J_6$ in zonal harmonics, plus the 
gravitational perturbations of
the five large moons close to the region, namely Janus, Epimetheus,
Mimas, Enceladus, and Tethys, in the first set. For the second set
we also added the gravitational perturbations from Aegaeon, 
Methone, Anthe, and Pallene.
To account for the rapid
orbital oscillations due to the oblateness
of Saturn, the state vectors resulting from the integrations are 
everywhere transformed to 
geometric elements following \citet{Renner06}.\\ 

In order to construct the frequency analysis map we explore the evolution
of the quantity $z'(t)=a(t)\exp(i\lambda)$, for each particle, 
where $a$ is the geometric 
semimajor axis and $\lambda$ is the mean longitude of the orbit. We have 
$T\sim9$ yr, equivalent to half the total 
integration time. We applied the FMFT \citep{Sidli96} 
over $z'(t)$ on the adjacent time intervals, $[0,T]$ and $[T,2T]$, 
in order to compare the main frequencies of $z'(t)$, $\nu_1$ and
$\nu_2$, from each time interval.\\

Since we are 
interested in the global dynamical structure of the region, we focus on the 
diffusion parameter, $D$,
for each particle, defined following \citet{Robutel01} and 
\citet{Correia05}, as:

\begin{equation} 
D=\frac{\left|\nu_1-\nu_2\right|}{T}.
\label{Eq:diffp}
\end{equation}

The $D$ parameter provides a measure of the particle's orbit
stability. Some particles do not survive the whole simulation and
we do not compute $D$ for them. For unstable particles, 
that nonetheless survive until the end of the simulation, 
the difference $\left|\nu_1-\nu_2\right|$ will be significant,
leading to large values of $D$. On the contrary, for stable particles
the same difference will be small, leading to very small values of $D$.
An estimation
of a diffusion time-scale, $t_{Diff}$, this is, an estimation
of the time required until an appreciable change in the $n$ of the 
particle is
observed, or the time required for
a diffusion of the orbit in the radial direction, can be crudely 
estimated as
$t_{Diff}=(DP)^{-1}$ yr, where $P$ is the period of the orbit in yr.\\

The FMFT algorithm let us to measure with confidence the variations
of $\nu_1$ and $\nu_2$, in the very short-time integrations 
performed. In order to characterize the dynamical region of
interest, we have plotted, in a color scale, the $\log(D)$ for 
each particle, according to its initial position in the geometric 
$a$ vs $e$ phase-space plane. The 
resulting map of the entire region is shown in Fig. \ref{fig:FReg5bb}.\\

To highlight the relevance of the small moons 
over their environment, an analogous map to that of Fig. \ref{fig:FReg5bb}
is shown in Fig. 
\ref{fig:FRegFull}. This map was obtained from the integration of
the same set of test particles subject to the gravitational
perturbations from the five large moons, plus the four
small moons: Aegaeon, Methone, Anthe, Pallene, and an oblate Saturn.\\

The maps presented in Figs. \ref{fig:FReg5bb} and \ref{fig:FRegFull}, 
show the general dynamical characteristics of the 
region. In both figures, black lines delimit the 
position of constant pericenter 
and apocenter of test particles, at the positions of the apocenter
and pericenter
of each major body, respectively. In Fig. \ref{fig:FReg5bb}, the 
positions of the four large moons are pointed by
filled black circles. Near the coorbital moons 
Janus-Epimetheus, the constant pericenter and apocenter lines are 
those corresponding to 
the position of Janus' apocenter and pericenter, respectively, as it
is the more massive moon of the two coorbitals. For Mimas,
both curves of constant apocenter and pericenter are plotted. For 
Enceladus, only the line delimiting the constant apocenter of particles at
the position of Enceladus' pericenter is shown. 
The redder colors in both maps are enclosed by such black lines. 
Particles that enter, or that are initially located inside such 
regions, are strongly perturbed
by the major bodies, resulting in a major probability for them to
collide with the corresponding moon, thus being lost from the simulation. 
In the map of Fig. \ref{fig:FReg5bb} most particles inside 
the Enceladus region are removed before 
the end of the simulation, while much less 
particles are removed inside the regions of Mimas and 
Janus-Epimetheus. However, some of the particles inside the Enceladus region
survived, remarkably those that are coorbital with the large moon.
A small patch of stability is also seen for particles of
small eccentricity, coorbital with Mimas, while no evident 
coorbital stability region is seen inside the
Janus-Epimetheus region. Outside of the regions delimited by black
lines, the more stable particles are those of small eccentricity,
expected since they do not cross the orbit of any large moon.\\

Regarding particle collisions with the large moons we
see that, for the simulation shown in Fig. \ref{fig:FReg5bb}, 
due to the Mimas larger eccentricity, compared to
that of Enceladus or Janus and Epimetheus, it takes 
longer time for Mimas to clean up its region despite its large size. 
From the $1\,454$ particles initially
inside the Mimas region, only $\sim$12\% of them collided with the 
large moon.
The surviving particles have an average $Log(D)=-1.4091149$.
This implies a diffusion time of
$t_{Diff}=9.946\times10^3$ yr, considering the orbital period of Mimas.
For Enceladus we found 885 particles initially inside
its region, out of which almost 70\% collided, 
and the surviving particles are mainly coorbitals
with the large moon. The surviving particles have an average
$Log(D)=-3.2606267$, which implies a
$t_{Diff}=4.857\times10^5$ yr, considering the period of Enceladus. 
Finally, in the region dominated by
Janus-Epimetheus, out of the 806 particles initially inside 
the corresponding region, around 36\% collided with any of the large
moons. The average $Log(D)=-1.5613890$ of the surviving
particles leads to a diffusion time of $t_{Diff}=1.915\times10^4$ yr,
by considering the period of Janus.\\

In the simulation shown in Fig. \ref{fig:FRegFull}, 
Aegaeon, Methone, Anthe, and Pallene were 
included as n-bodies. The 
positions of the four 
small moons are indicated by brown circles in the map. All lines for 
constant pericenter and apocenter of test particles at the positions
of apocenters and pericenters of the large and small moons are shown in
black solid lines. In this case, a greater number of 
particles is removed
from the collision regions (those enclosed by solid black lines) 
leading to the appearance of big white zones. As 
in the previous case,
inside these collision regions, some coorbital particles survive, 
mainly with Enceladus and with Mimas to a lesser extent.\\

The results from the map of Fig. \ref{fig:FRegFull} suggest 
that the small eccentricity of Aegaeon, Methone, and Anthe, let them to 
better clean up their orbits in a short period of time 
($\sim$18 yr), when compared to the much more
massive Janus, Epimetheus, or Mimas. Nonetheless, inclinations
likely play also an important role for the clearing ability of the 
moons. To test this, we perform a simulation with the same conditions 
of that of Fig. \ref{fig:FRegFull}, but this time with the 
initial inclinations of all the 
nine moons set to zero. The result is shown in the map of Fig.
\ref{fig:FRegZero}. Clearly, when the moon lies in the orbital
plane, it better clear its path, this is manifest in the large white
regions inside the black lines limiting the regions of Janus-Epimetheus and
Mimas. Moreover, the coorbital stability region is clearly visible now
for Mimas, and two high eccentricity coorbitals with Janus-Epimetheus
also survive. In the case of Aegaeon, Methone, and Anthe, the changes are
insignificant as all they had from the beggining a small inclination 
($<0.02^\circ$). The same lack of change is observed for Enceladus, 
which has an $i\approx0.004^\circ$.\\

Therefore, moons of small eccentricity and inclination
clean their orbits efficiently, regardless their mass (see Aegaeon, 
Methone, Anthe, and Enceladus), and coorbital particles with the  
lowly inclined moons are more likely to survive. On the other hand, 
moons with larger eccentricities and inclinations require longer 
times for the clearing of their orbits despite their large masses
(see Janus-Epimetheus, Mimas, and Pallene), and coorbital particles 
are not likely to coexist with them due to a large inclination. 
Interestingly for Pallene, from the previous exercises we 
have gained an important hindsight about the high stability of the region
surrounding this small moon. The combination of a small mass and
a large inclination ($\sim0.182^\circ$) make Pallene a 
weak perturber and inefficient clearer of its region.\\

In order to explore in detail the dynamical environment of the
small moons, we mapped the geometric phase-space around them, 
zooming from the general view
into three finer maps, one around Aegaeon, a second around
Methone and Anthe, and one more around Pallene. Although
the simulations last the same amount of time as the ones describe
earlier, the coverage of $a$ and $e$ is far better. We describe the 
initial conditions used for each map in the next subsections.\\

\subsection{The Dynamical Environment of Aegaeon}

\begin{figure*}
\includegraphics[width=17cm]{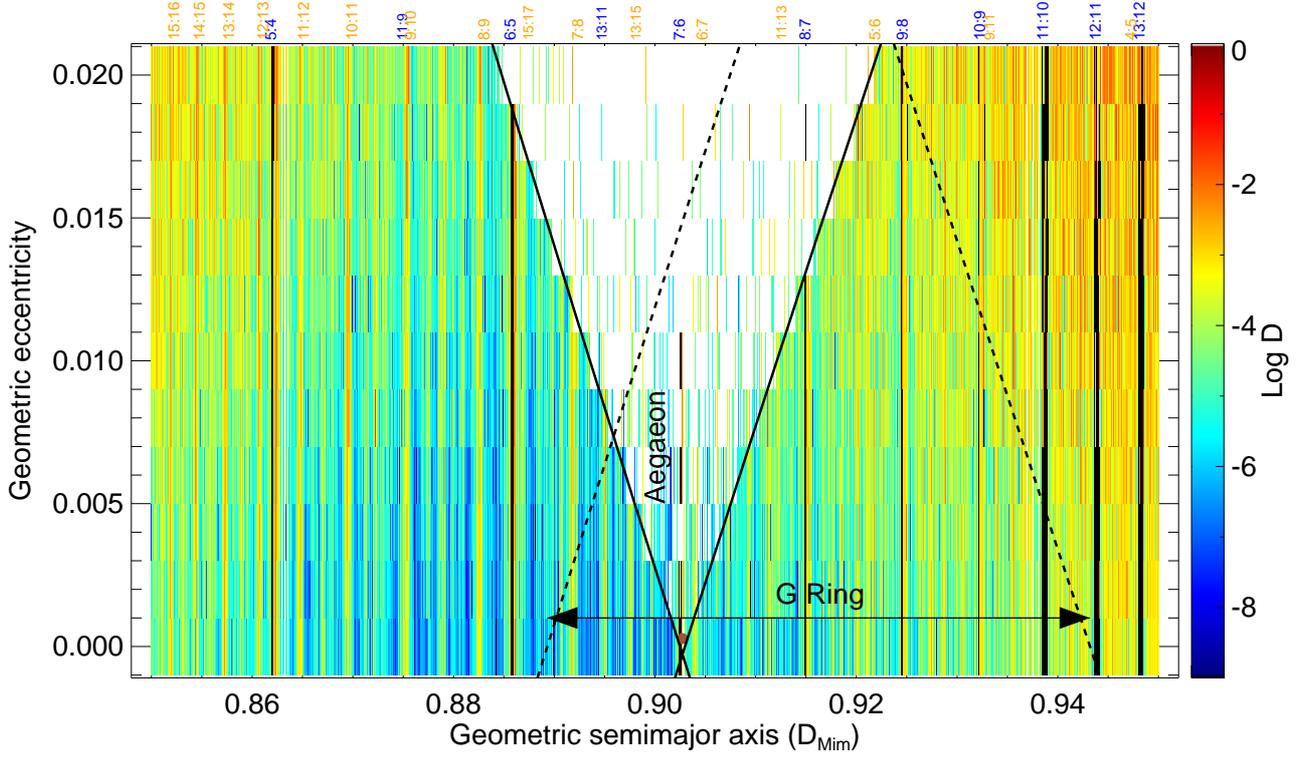}
\caption{Diffusion map for the Aegaeon region. The location
of Aegaeon is indicated by a brown circle. Solid black lines
stand for the pericentric and apocentric distances of particles
at the apocenter and pericenter distances of Aegaeon, respectively.
Most particles 
inside this region collide with Aegaeon; however, some particles 
survived, mainly those with the smaller eccentricity. 
Resonant particles in 
first order MMRs with Mimas are colored in black. Labels for first and
second MMR ratios are shown at 
the top of the figure, to indicate the location of the respective resonance.
Blue labels are for resonances with Mimas, orange ones 
for resonances with Janus.
Black dashed lines mark the position of constant pericentric and
apocentric distances of particles, at the position of the inner and 
outer G ring edges, respectively. All particles 
inside this region are assumed to form part of the G ring.}
\label{fig:AReg}
\end{figure*}

\begin{figure}
\includegraphics[width=\columnwidth]{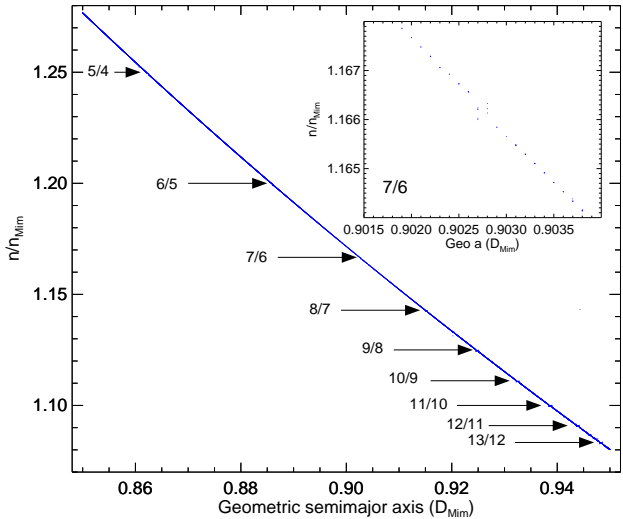}
\caption{Frequency curve for the Aegaeon region. We plot 
the ratio of particle's mean
motion, $n$, over Mimas mean motion, $n_{Mim}$, \textit{vs}
initial geometric semimajor axis of the particle. Plateaus in the curve
correspond to the location of MMRs with Mimas. We highlight one 
of such plateaus,
for the 7:6 MMR, close to the 7:6 CER where Aegaeon librates, to
exemplify the flattening of the frequency curve.}
\label{fig:ARegRes}
\end{figure}

To get a better understanding of the dynamical evolution of Aegaeon and
particles nearby, like the ones forming the G ring, 
we perform the numerical integration of the orbits  
of $11\,000$ test particles, covering a small
patch of the geometric $a$ vs $e$ phase-space around the small moon. 
In $a$ we cover
from 0.85 to 0.95 D$_{Mim}$, in
steps of $1.0\times10^{-4}$. In $e$ we cover from 
0.0 to 0.02 in steps of $2\times10^{-3}$. Initial inclination 
and the angles $\bar{\omega}$, $\Omega$, and $\lambda$, 
are all set to zero. Test particles are subject to gravitational  
perturbations from the oblateness
of Saturn up to $J_6$, the five large moons of 
the global case described earlier, and  
Aegaeon, included as a massive body of $1.89\times10^{13}$ g, assuming
a bulk density of $0.54$ g/cm$^3$ \citep{Thomas13}. The 
resulting diffusion map is showed in Fig. \ref{fig:AReg}.\\

The average
location of Aegaeon is shown in the map of Fig. \ref{fig:AReg} 
by a filled brown circle. This location is close but not
equal to the location of the 7:6 MMR, as Aegaeon librates in the 7:6 
CER with Mimas. The
long-term dynamics of Aegaeon in this resonance will be further analyzed
in section \ref{sec:longt}. In this map, black solid lines
indicate the constant apocenter and pericenter distances of particles, 
at the position of Aegaeon's pericenter and apocenter, respectively. Most 
of the particles inside this region end up colliding with the small
moon before the end of the simulation, which results in a wide 
white triangle feature. Initially, $2\,042$ particles were inside 
the region enclosed by solid black lines, while just 303 of them survived, 
i.e. almost 85\% of the
particles collided with Aegaeon in just $\sim$18 yr. In average, surviving
particles have $Log(D)=-4.6857625$, which implies a 
$t_{Diff}=2.192\times10^7$ yr.\\

As we can see from Fig. \ref{fig:AllRes}, a great number of
first and second order MMRs, mainly with Janus and Mimas, populate 
the region around Aegaeon. The evenly colored vertical 
features in the map of Fig. \ref{fig:AReg}, both interior and 
exterior to the orbit of Aegaeon, are related to any of these MMRs.
To identify the strongest resonances, and therefore particles
librating inside them, we make use of the frequency curve
\citep{Robutel01}. In Fig \ref{fig:ARegRes}
we show the frequency curve, i.e. the ratio of the particle's mean 
motion over Mimas mean motion, $n/n_{Mim}$, \textit{vs} the initial 
particle geometric $a$. MMRs are characterized for the flatness of
the frequency curve, due to a constant $n/n_{Mim}$ ration of nearby
particles of different $a$; this facilitates the identification
of librating particles. In the curve 
of Fig. \ref{fig:ARegRes}, we point out (with arrows in the figure) 
to the presence of the next first order MMRs: 6:5, 7:6, 8:7, 9:8, 
10:9, 11:10, 11:12, and 
13:12; although the flatness of the frequency curve associated to the
mentioned resonances is barely visible,
some particles cause the curve to flatten, making possible their 
identification.
To exemplify the way how frequency curve flats in resonances, 
in Fig. \ref{fig:ARegRes} we zoomed into a region
very close to the 7:6 MMR with Mimas, where
a number of particles maintains a constant ratio $n/n_{Mim}=7/6$.
We note that the small maximum eccentricity covered by our grid of
initial conditions
accounts for the width of the resonant region to be very small, thus
making difficult for such regions to stand out from the Keplerian trend
of the rest of the frequency curve.\\ 

It is in this sense that the resonant regions are better 
recognized directly from the diffusion map of Fig. \ref{fig:AReg}. 
In the map we have plotted as black rectangles the resonant particles with
Mimas identified from the frequency curve, i.e. the particles for 
which the ratio $n/n_{Mim}$ is equal to the corresponding MMR ratio $p/q$.
Nonetheless, apart from this first order MMRs with Mimas, other 
homogenous vertical regions are clearly recognizable, and other 
resonances can be related to them. For this, we have indicated at 
the top of the
Figure the corresponding ratios at the position of first and
second order MMRs with Janus, in orange, and with Mimas, in blue. 
Such resonances could not be as strong as first order MMRs with Mimas,
however, their influence is imprint in the homogeneous $D$ parameter
they produced over particles of different eccentricities.\\

We highlight the importance
of MMRs on the evolution of test particles, even as there are
not known structures or particles in such regions in the real system, 
i.e. outside the edges of the G ring. 
If ever a population of particles inhabited some of the stable regions 
currently empty,
then it needs to be an explanation, other than the pure gravitational 
evolution, for the lost of such particles. Non-gravitational forces
are a likely explanation, given that particles inhabiting 
such regions are expected to be small and therefore subject to 
drift forces like solar radiation pressure and plasma drag 
\citep[see for example][Madeira et al., in prep.]{Sun17}.\\

Regarding the G ring, we know it is formed by $\mu$m-sized 
particles and extends from 
$\sim165\,000$ to $175\,000$ km from Saturn centre, or from 
$\sim$0.8893 to 0.9432 D$_{Mim}$ \citep{Horanyi09}. In Fig. 
\ref{fig:AReg} the location 
and size of the G ring are indicated by the
double-headed black arrow. Also showed in the figure, in dashed lines,
are the constant pericenter and apocenter of particles at the position of 
the inner and outer edges of the G ring, respectively. It is assumed
that particles inside this region can form part of the ring, since
particles outside of it (some of them with the same $a$ but larger $e$) 
would have pericentric and apocentric excursions farther away from 
the observed edges of the ring.
Initially $3\,913$ particles form 
part of the ring, while $2\,937$ survive at the end of the 
simulation. This is, around
25\% of the G ring particles are lost due to collisions with
Aegaeon. Besides, surviving particles have in average a 
$Log(D)=-4.3779971$, which implies a $t_{Diff}=1.079\times10^7$ yr.
Such particles are 
dynamically heated, i.e. by increasing their inclination, from 0 to 
0.00425$^\circ$ in average, mainly due to the effect of MMRs with Mimas.
This average inclination implies that the G ring is vertically widened 
in a short time period, by $\sim12.6$ km, considering an average 
$a\approx170\,000$ km
for the vertical excursions of particles ($a\times i$).\\

\subsection{The Region Around Methone and Anthe}

\begin{figure*}
\includegraphics[width=17cm]{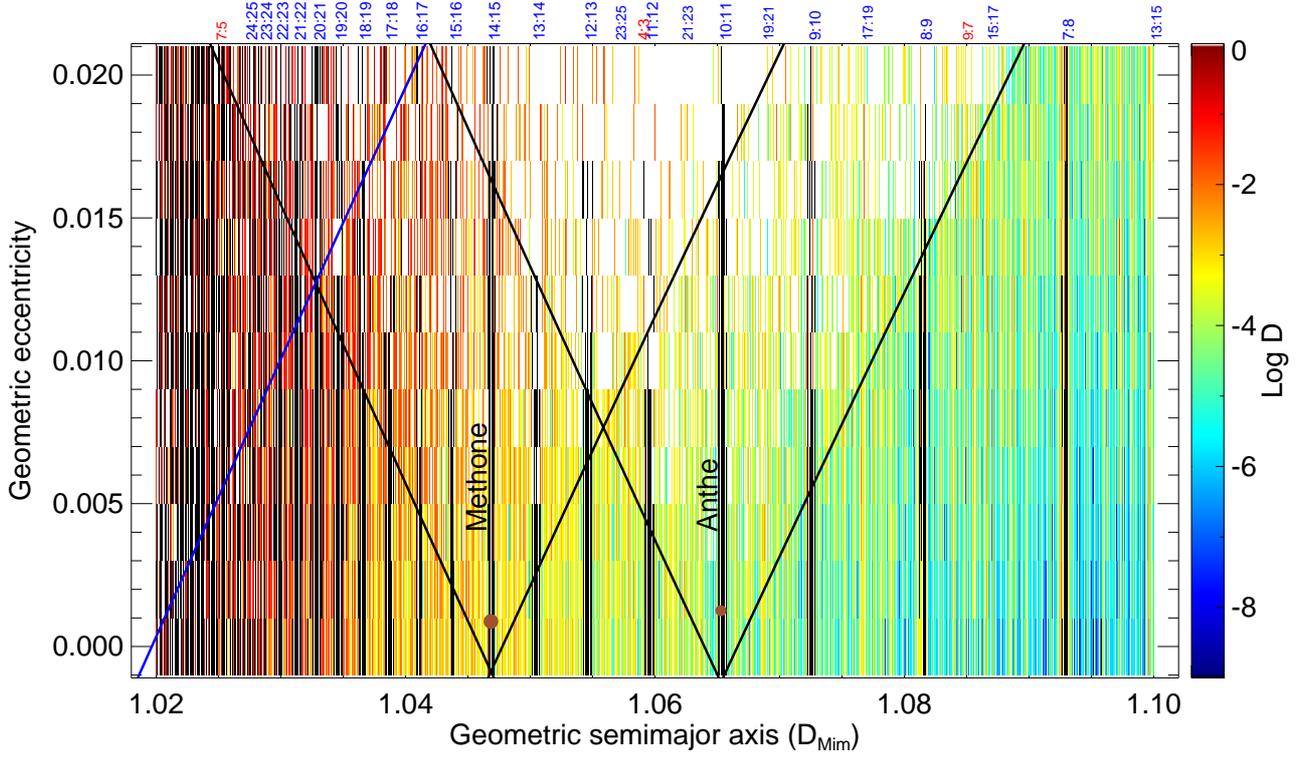}
\caption{Diffusion map for the region around Methone and Anthe. 
The locations of the small moons are indicated by filled brown circles. 
Solid black lines stand for the apocentric and pericentric distances of
particles at the position of pericenter and apocenter of the
small moons, while the blue line is for the pericentric distance
of particles at the apocenter of Mimas. 
As in Fig. \ref{fig:ARegRes}, particles in first order MMRs with Mimas 
are colored black; closer to Mimas (to the left of the map) 
such resonances are tightly packaged. The resonant ratios for first 
and second order MMRs with Mimas (in blue) and Enceladus (in red), are
shown at the top of the figure. An important percentage of the
map surface is unstable due to the  
overlapping of collision regions of Mimas, Methone, and Anthe.}
\label{fig:MReg}
\end{figure*}

\begin{figure}
\includegraphics[width=\columnwidth]{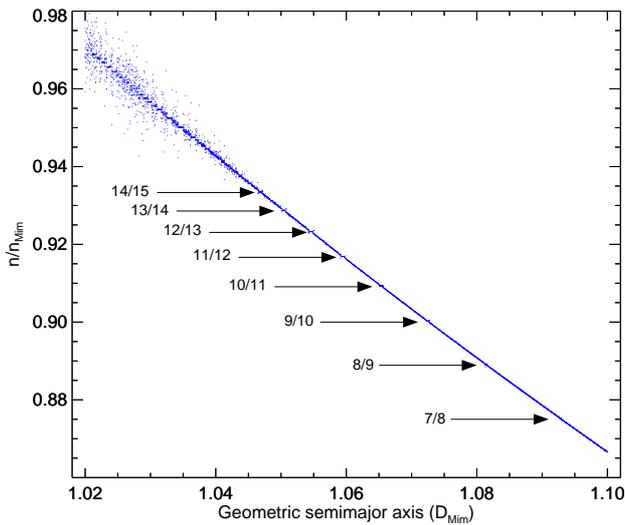}
\caption{Frequency curve for the Methone and Anthe region.
Same as in Fig. \ref{fig:ARegRes}, we point to the plateaus of MMRs
with Mimas. In this case, closer to Mimas (top-left in the figure)
the unstable region is easily recognizable by the dispersion of the 
ratios.}
\label{fig:ARegMet}
\end{figure}

For Methone and Anthe, we study the region around
both moons in the same map, due to their proximity. The
simulation performed for this aim covers from 1.02 to 1.10 D$_{Mim}$
in geometric $a$ and from 0 to 0.02 in geometric $e$. 
The steps in $a$ and $e$ are, 
as in the previous case, $1.0\times10^{-4}$ and $2\times10^{-3}$,
respectively. This makes a total of $8\,800$ test particles.
Inclination is set to 0 as well as $\bar{\omega}$, $\Omega$, and $\lambda$.
Particles are subject to the oblateness of Saturn, gravitational
perturbations from the five large moons, plus Methone and Anthe, whose
masses were considered to be: $9.38\times10^{14}$ g for Methone and
$3.5\times10^{14}$ g for Anthe. Regarding densities, we assumed 
values of 0.31 g/cm$^3$ for Methone, following \citet{Thomas13},
while for Anthe we used a value of 0.35 g/cm$^3$,
by considering an average of the densities of the other three small moons.\\

The diffusion map for the region of Methone and Anthe is shown in
Fig. \ref{fig:MReg}. Brown filled circles mark the average position 
of the small moons. Black lines indicate the pericentric and 
apocentric distances
of particles at the positions of the apocenters and pericenters of 
Methone and Anthe, respectively. The blue line mark the pericentric
distance of particles at the apocenter of Mimas.\\

An important fraction of this map 
is unstable, mainly due to the overlapping of the collision
regions of Mimas, Methone, and Anthe. Nonetheless, some particles
in the region interior to the orbit of Methone, lie in MMRs with Mimas, 
which are tightly packaged the closer they are
to the large moon. Another stability zone exists, located 
between the Methone and Anthe locations, while the predominantly 
stable region can be found 
beyond the orbit of Anthe, where the perturbative influence of Mimas is 
weaker. 
Inside the combined collision regions of Methone and Anthe, there
are initially $4\,204$ particles, out of which $2\,254$ survive the 
whole simulation, therefore, around 46\% of the particles end up
colliding with any of the moons. For the surviving particles an average 
$Log(D)=-2.9265259$ is found, which implies a $t_{Diff}=3.023\times10^5$
yr. This is a considerable shorter stability time than that found for
the surviving particles inside the Aegaeon region; this since although
there are few of them, several
of the Aegaeon surviving particles are those of smaller eccentricity,
while others remain in resonance, 
avoiding larger perturbations from Mimas and collision with Aegaeon.
In this case, however, many particles that survive the simulation
are part of the high eccentricity and largely perturbed region of
the map, for which it is to be expected shorter diffusion time-scales.\\

Fig. \ref{fig:ARegMet} shows the frequency curve for the 
Methone/Anthe region. Interestingly, the unstable region close to
Mimas is very easy to identify due to the scatter of the points 
(or $n/n_{Mim}$ values) away of the Keplerian trend. In the figure we
point out the existence of the Mimas MMRs: 7:8, 8:9, 9:10, 10:11, 
11:12, 12:13, 13:14, and 14:15. The particles with the constant
ratios responsible for the plateaus are colored in black in the
map of Fig. \ref{fig:MReg}.

\subsection{The Dynamical Environment of Pallene}
\label{sec:palreg}

\begin{figure*}
\includegraphics[width=17cm]{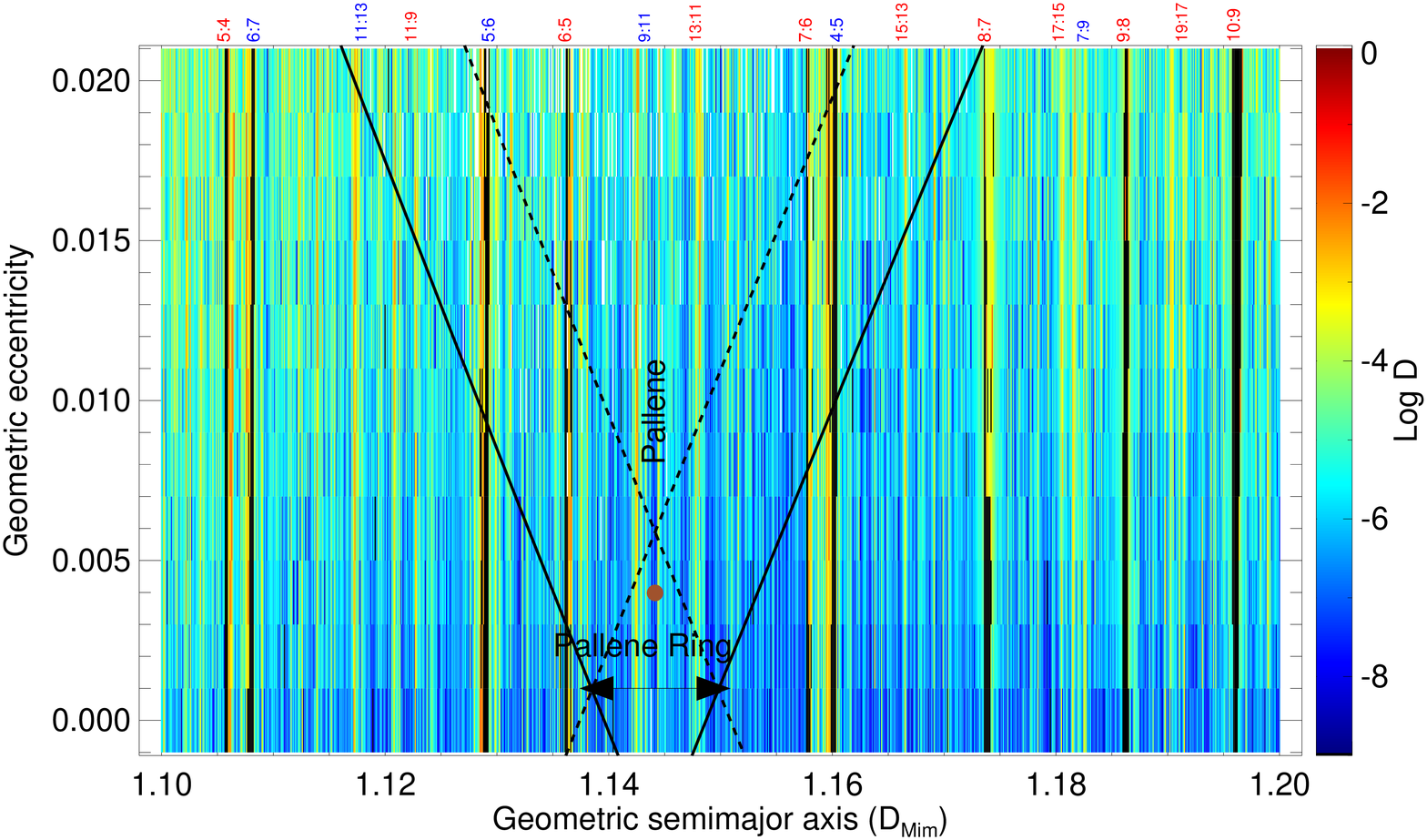}
\caption{Diffusion map for the region around Pallene. The 
average location of Pallene is indicated by a filled brown circle. Solid 
black lines stand for the pericentric and apocentric distances of 
particles at the apocenter and pericenter distances of Pallene, 
respectively. Black dashed
lines mark the position of constant pericentric and apocentric distances 
of particles, at the position of the inner and outer edges of the 
Pallene ring, respectively. The width and location of the ring is
accounted by the double-headed arrow. All particles inside the region
enclosed by dashed lines are 
assumed to form part of the ring. Particles in first order MMRs 
either with Mimas or Enceladus are colored in black. 
The first and second order MMR 
ratios are shown at the top of the figure, in blue for resonances
with Mimas and in red for resonances with Enceladus. Due to 
its small size and large $i$, Pallene is unable to clean up
its orbit in only $\sim18$ yr, or approximately $5\,700$ of its
orbital revolutions.}
\label{fig:PReg}
\end{figure*}

The region around Pallene was studied through a simulation
of test particles covering a patch of geometric $a$ vs $e$ phase-space,
going from 1.1 to 1.2 D$_{Mim}$ in $a$, in steps of $1.0\times10^{-4}$. 
In $e$ the values run from 
0.0 to 0.02, in steps of $2\times10^{-3}$, totalizing 
$11\,000$ test particles. They are subject to the gravitational 
perturbations from an oblate Saturn,
the five large moons of the region plus Pallene. The mass of 
Pallene is taken to be $2.69\times10^{15}$ g, with a density of
0.25 g/cm$^3$ \citep{Thomas13}.\\

The diffusion map for the Pallene region is shown in Fig. \ref{fig:PReg}.
Same as in previous cases, the average location of Pallene is mark by the
filled brown circle. Solid black lines delimit the region
where particles have the highest 
probability for colliding with Pallene. Dashed black
lines delimit the apocentric and pericentric distances of particles
at the position of the inner and outer edges of
the Pallene ring, respectively. The width of the ring is 
accounted for with the 
double-headed arrow. The ring is about $2\,500$ km in width and
centered at the small moon \citep{Hedman09}. Interestingly, the 
radial extend of the ring is constrained by the largest pericentric and
apocentric excursions of Pallene itself.\\

We can see from Fig. \ref{fig:PReg} that Pallene does not lie close to 
any first order MMR, neither with Mimas or Enceladus, 
suggesting the non-resonant dynamics of the small moon and its 
ring. If by chance, the second order MMRs 9:11 with Mimas and
13:11 with Enceladus, lie 
close to the position of Pallene, these weak resonances seems 
unable to affect the dynamics of the small moon whatsoever.
Nonetheless, a rich structure of resonances shapes this intermediate
region between the major satellites Mimas and Enceladus, as can
be seen from Fig. \ref{fig:AllRes}. In
the map of Fig. \ref{fig:PReg} we plot in black the resonant
particles in first order MMRs either with Mimas or Enceladus. At the top
of the figure we indicate the ratios of the resonances (in blue for
Mimas, in red for Enceladus). It is clear that some of
the vertical features visible on the map are not related to first
order MMRs, therefore we also indicate, by labeling the ratio at the top
of the figure, the location of second order MMRs with Mimas 
and with Enceladus, with the same color code. Despite the abundance of such 
first and second order MMRs, all the region is globally very stable and
only a minor fraction of particles are lost in our short time 
integration.\\

Inside the collision region of Pallene, there are initially
$3\,522$ particles, while $3\,375$ of them survive at the end 
of the simulation. This is, less than $\sim$5\% of the particles
collide with Pallene in $\sim$18 yr. In average,
for the surviving particles we have a $Log(D)=-5.4219$, which
implies a stability time of $t_{Diff}=8.365\times10^7$ yr, the longest
for any of the regions studied so far. Regarding the Pallene ring, 
there are initially
inside the region delimited by the dashed lines a total 
of $266$ particles, out of which only one collided with Pallene. A $Log(D)=-6.0716074$ is
found in average for the ring surviving particles, leading
to $t_{Diff}=3.733\times10^8$ yr. The
average inclination of the surviving ring particles is 
0.001023$^\circ$. This inclination implies vertical
excursions of just $\sim3.8$ km, considering an
average $a\approx212\,000$ km. This is an order of magnitude smaller
than the vertical width of around 50 km, determined by \citet{Hedman09} 
for the Pallene ring. Nonetheless, we expect a larger value of
the final average inclination in longer simulations. Such explorations
are planned to be presented in a future work.\\ 

It is worth to note the high stability of the orbits inside the region of
the maximum orbital excursions of Pallene. Even more, particles of small
eccentricity laying close to the small moon, at least below
$e\sim0.006$, are those actually forming
part of the ring and are expected to survive in a long-term basis. 
If the ring itself is a stable feature, it should then be formed by
this low $e$ particles, leading in its way to a ring of small 
eccentricity also.
Due to the limited number of Cassini images in which the Pallene ring 
is visible, it has not been possible to accurately determine the 
eccentricity of the ring, among other properties such as its mass.
Nonetheless, the ring seems to follow closely the orbital path
of the small moon. This fact implies that most likely the ring 
shares the same orbital
properties of the parent moon. A further detailed study about the ring 
is planned to be presented in a future paper.\\

\section{Long-term evolution of Aegaeon, Methone, Anthe, and Pallene}
\label{sec:longt}

\begin{figure}
\includegraphics[width=\columnwidth]{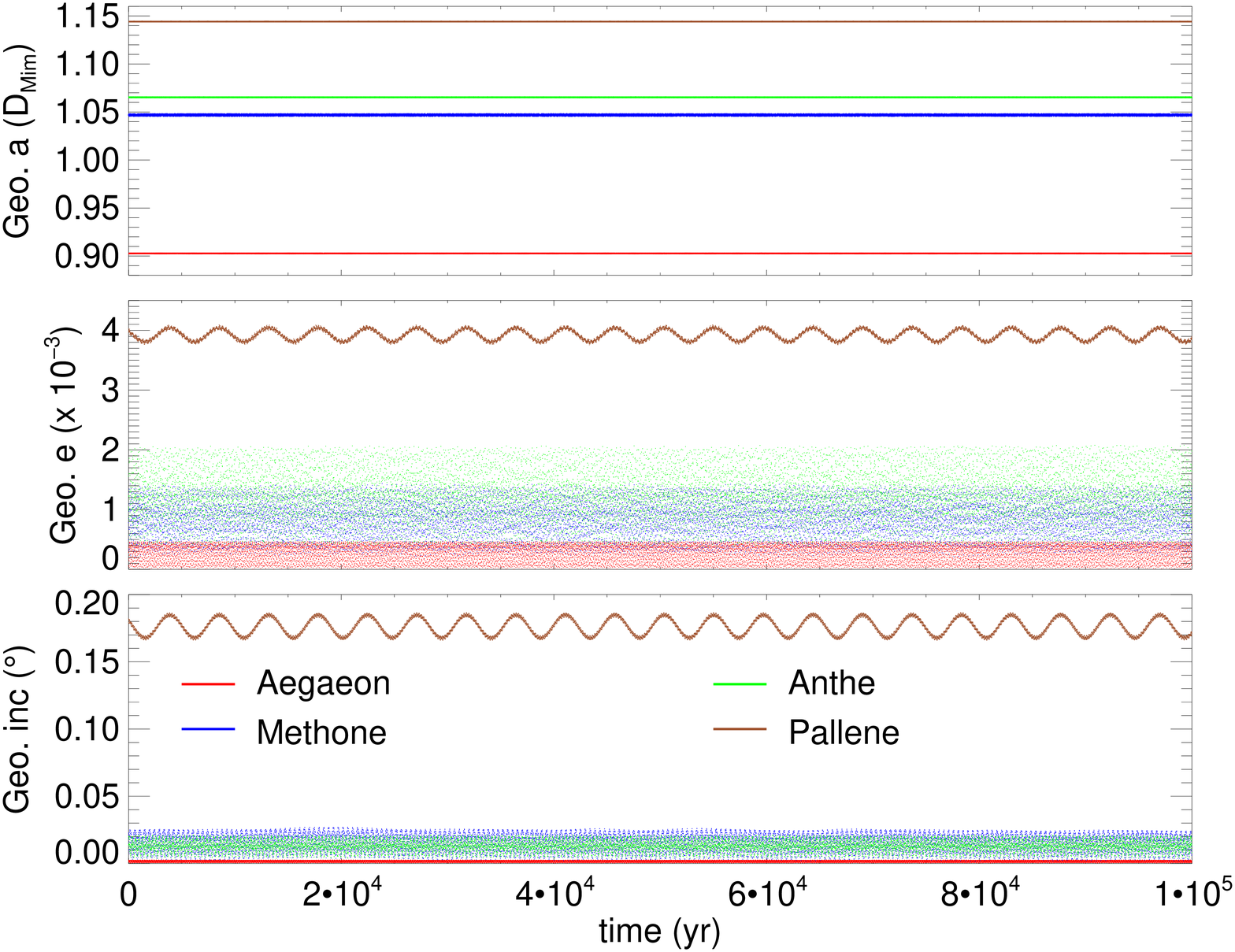}
\caption{Long-term evolution of geometric orbital elements for
Aegaeon, Methone, Anthe, and Pallene. Variations in geometric semimajor
axis (top panel) are hardly visible at this scale. Middle panel shows
the evolution of geometric eccentricity, which larger variations are
for Anthe (in green) and remarkably periodic for Pallene (brown).
Bottom panel shows the evolution of geometric inclination. Pallene
shows the larger variations, again periodic, besides the larger
inclination among the small moons.}
\label{fig:ltevol}
\end{figure}

\begin{figure}
\includegraphics[width=\columnwidth]{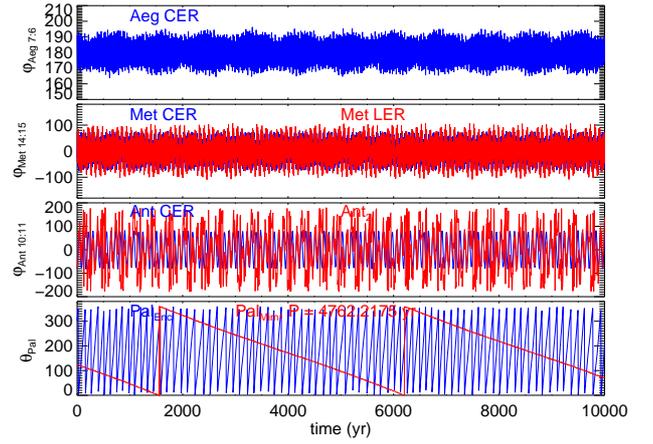}
\caption{Long-term evolution of different resonant arguments 
of the four small moons. We show the same arguments as the ones 
presented in Fig. \ref{fig:reso} but now for the initial 
$10\,000$ yr of the long-term simulation. All arguments for CER
with Mimas (blue curves in three first panels) remain librating 
for Aegaeon, Methone and Anthe for the whole simulation time, so the moons
remain in resonance. Pallene, on the other hand, is never in
resonance. Nonetheless, the quasi resonance with Mimas suggested by 
\citet{Callegari10} seems to perturb
its orbit, inducing periodic variations in $e$ and $i$.}
\label{fig:reslte}
\end{figure}

\begin{table*}
\centering
\caption{Summary of mean values for the evolution of geometrical
orbital elements and frequency analysis.}
\label{tab:Tablt}
\begin{tabular}{@{}lcccccc}
\hline
\hline
Satellite & $\bar{a}_{Geo}$ (D$_{Mim}$) & $\bar{a}_{Geo}$ (km) & 
$\bar{e}_{Geo}\times10^{-3}$ & $\bar{i}_{Geo}\times10^{-2}$ ($^\circ$) & 
$\log(D)$ & $t_{Diff}$ (Myr) \\
\hline
Aegaeon & $0.90272626\pm3.472\times10^{-5}$ & $167\,490.93\pm6.44$ & 
$0.27542\pm0.23335$ & $0.16156\pm0.12165$ & 
-3.0138 & $0.466565$ \\
Methone & $1.0468417\pm1.531\times10^{-4}$ & $194\,229.97\pm28.40$ &
$0.85436\pm0.61407$ & $1.5441\pm1.2678$ & 
-3.1172 & $0.473853$ \\
Anthe & $1.0652882\pm1.514\times10^{-4}$ & $197\,652.51\pm28.09$ & 
$1.2091\pm0.8610$ & $1.3255\pm0.8787$ & 
-3.1291 & $0.474469$ \\
Pallene & $1.1441441\pm1.760\times10^{-5}$ & $212\,283.35\pm3.26$ & 
$3.9240\pm0.1677$ & $17.6289\pm1.0308$ & 
-5.3085 & $64.41534$ \\
\hline
\end{tabular}
\medskip
\end{table*}

In this section we present the results of a single 
long-term simulation, lasting for $10^5$ yr, 
which included the nine large bodies of Table \ref{tab:tabpar}, and the 
oblate Saturn with physical parameters given in Table \ref{tab:tabSat}. 
In Fig. \ref{fig:ltevol} we plot the evolution of the geometric orbital
elements of the four small moons of Table \ref{tab:Tabinism}. 
For them, we calculate 
the mean and maximum variations of the geometric parameters. 
Also, we apply the frequency analysis to 
the small moons in order to estimate a diffusion parameter and 
an associated diffusion time-scale for them in the current 
configuration of the saturnian system. 
Our results are summarize in Table \ref{tab:Tablt}.\\

Second and third columns of Table \ref{tab:Tablt} show the 
mean and maximum variations of the 
geometric semimajor axis, $\bar{a}_{Geo}$, in units of D$_{Mim}$ 
(second column) and
km (third column) for an easier interpretation. 
The largest variations in this parameter are
those of Methone and Anthe. We can see that all
the orbits of the small
moons are highly stable over the $10^5$ yr simulation. The largest 
ones, for Methone and Anthe, are of only $\sim28$ km. 
It has been suggested
that those two moons are closer to a chaotic zone produced by the 
overlapping of the CER and LER resonances with Mimas 
\citep{Elmoutamid14}. The variations, well above the 4 km separation of
the CER and LER may lead to Methone and Anthe to be influenced by both
resonances at different times, nonetheless, the combined perturbations
are weak enough as being ineffective in leading to a chaotic evolution
of their orbits in $10^5$ yr.\\

In the case of Aegaeon, the variations $<6.5$ km implies that
the moon is well trapped inside the 7:6 CER with Mimas. For 
Pallene, the variations of only $\sim$3 km turns it into the more stable
of the small moons. Pallene is far from any first order
MMR, this allows it to avoid any strong perturbation from Mimas
or Enceladus.\\

Regarding geometric $e$ and $i$, Anthe turns out to be the 
more perturbed in $\bar{e}_{Geo}$, probably as a result of 
perturbations produced by the overlapping the CER and LER with Mimas, 
being that, as \citet{Elmoutamid14} have pointed out, Anthe is  
closer to the chaotic zone than any of the other small moons, although
variations in this parameter for Methone are comparable.
Regarding inclinations, for Aegaeon, Methone, 
and Anthe, they are also small, so the small moons remain very
close to Saturn's equatorial plane; Aegaeon reachs maximum
excursions above the plane of only $\sim4.7$ km, while for Methone
and Anthe vertical excursions are of $\sim52$ and $\sim46$ km,
respectively. For Pallene such excursions reach $\sim653$ km.\\

The orbital behavior of Pallene is interesting, showing that 
variations in $e$ and $i$ are 
periodic and clearly correlated. The period of such oscillations is
$4\,762.21$ yr, for both $e$ and $i$, and although its mean 
values of $e$ and $i$ are the largest among the small moons, 
it presents the more stable orbit. 
We can confirm the last statement by calculating
the diffusion parameter and the associated diffusion time for the
orbits of the small moons, as shown in the last two columns of Table 
\ref{tab:Tablt}. The value of $D$ is determined from the evolution of
the same quantity $z'(t)=a(t)exp(i\lambda)$ as for the test 
particles of the diffusion maps.
Here, $t_{Diff}$ gives us an estimation of the time required for an
appreciable change in the orbital evolution of the small moons to be 
observed.
As shown in the last column of Table \ref{tab:Tablt}, Pallene could 
remain stable for as long
as $\sim$64 Myr in its current orbit, while Aegaeon, Methone, and Anthe 
could do the same for only $\sim0.45$ Myr. This short-time
period results from a crude estimation and could result from the 
large variations of the three small moons produced by the resonances
they are trapped in. Nonetheless, for Pallene it seems to represent
a confident value for the stability of the moon in a long term basis.\\

Regarding the resonant nature of the small moons, we show in Fig. 
\ref{fig:reslte} the long term evolution of the same resonant
arguments previously shown if Fig. \ref{fig:reso}, where
we explored their
short-term evolution. Now we show a period of $10\,000$ yr out
of the $10^5$ yr of our simulation, as the behavior remains
unchanged along the whole run. From Fig. \ref{fig:reslte}
we can conclude that the three small moons currently trapped in CERs with 
Mimas, will remain there for a long-term basis, as is shown by blue
curves in the three first panels of Fig. \ref{fig:reslte}. Such
arguments remain librating for the whole time of the simulation. 
Aegaeon, Methone, and Anthe are 
strongly affected by CER resonances and
they could hardly escape from them, probably explaining the existence
of their associated arcs. Indeed, if the small moons remain for a long 
time trapped in the 
resonances, the slow impact-ejecta processes that are believed to
suply the arcs with dusty material, could have plenty of time to act
and originate
such structures, even if the tiny grains are constantly removed
due to the action of gravitational and non-gravitational forces,
such as collisions, solar radiation, and plasma drag.\\

For Methone, the argument associated to the LER with Mimas also remains
librating for the whole simulation, therefore, this resonance 
also affects the dynamics 
of the small moon; this could explain the large variations in $a$ 
seen for Methone. In the case of Anthe, the argument 
$\varphi_2=11\lambda_{Ant}-10\lambda_{Mim}-\bar{\omega}_{Ant}
-\Omega_{Ant}+\Omega_{Mim}$, shown in Sec.
\ref{sec:shortt}, starts librating 
but then it changes and starts to circulate,
then librates again repeating the cycle with a constant period 
and sequence. Although the erratic regime changes from libration
to circulation are a signature of chaotic behavior, in this case, the 
regularity in the behavior of this argument seems not to lead to
the chaotic evolution of Anthe.\\

In the case of Pallene, we show the two arguments previously 
presented in Sec. \ref{sec:shortt}; now it
is clear that both arguments circulate. The resonant argument 
associated with Mimas, $\theta=\bar{\omega}_{Pal}-\bar{\omega}_{Mim}-
\Omega_{Mim}+\Omega_{Pal}$,
circulates with a period of $\sim4\,762.21$ yr; this is the same period as
the one found for the long-term oscillations of $e$ and $i$. Such correlation
implies that Mimas induces those variations on the orbit of the small moon,
despite leaving it in a highly stable orbit.\\

\section{Conclusions}
\label{sec:conc}

In this work we have explored the dynamical evolution of four
Saturnian small moons: Aegaeon, Methone, Anthe, and Pallene, as
well as characterized their dynamical environment by using thousands
of test particles surrounding them, by means of
short and long term simulations, going up to $10^5$ yr, this is,
longer than any previous numerical exploration. We have considered
the current configuration of the Saturnian system, including the
oblateness of Saturn up to $J_6$ and the five largest moons close to the
region of interest, namely Janus, Epimetheus, Mimas, Enceladus,
and Tethys. Through frequency analysis and the long term evolution
of the moons we have analyzed their stability and that of the
region surrounding them. The frequency analysis has
allowed us to 
obtain a global dynamical perspective of the region inhabited
by the small moons by using detailed but short time integrations. 
With it we gain an estimation of the stability times for particles in such 
regions and it also let us characterize the resonances involved in shaping
or perturbing such regions. We have also found that:
\begin{itemize}
\item all the four small moons are stable in a long-term 
basis. If the current
configuration of the system would remain unchanged, this is, if the
migration of the largest moons, mainly Mimas, is slow enough, then
the stability time for the small moons would be at least  
$\sim0.4$ Myr
for Aegaeon, Methone, and Anthe, and up to 64 Myr for Pallene.
\item Aegaeon remains trapped in the 7:6 CER with Mimas, with
maximum variations of only $\sim6.4$ km in $a$. Its $e$ remains small, and
variations in $i$ implies maximum excursions above Saturn's equatorial
plane of 4.7 km.
\item Methone and Anthe share similar stability times, as well as
variations of the same order in their orbital parameters. Methone
librates in both the 14:15 CER and LER with Mimas, while Anthe 
remains in the 10:11 CER with Mimas but it is also
close to the chaotic region produced by the overlapping of the CER
and LER. The influence of two resonances on those small moons likely 
lead to largest variations than those observed for Aegaeon. For 
both Methone and Anthe, variations
in $a$ are of $\sim$28 km, while their $i$ imply vertical 
excursions of 52 and 46 km, respectively.
\item Pallene is the most stable of the small moons analyzed. It remains
as non-resonant but, nonetheless, it suffers long-term perturbations
from Mimas, through the quasi-resonance found by \citet{Callegari10}. 
Such perturbations only affect the evolution of $e$ and $i$, inducing
long-term oscillations, while its $a$ changes by only $\sim$3.2 km.
\item Regarding the G ring, 25\% of ring particles collide
with Aegaeon, while the remaining are stirred in their $i$ up to
$0.00425^\circ$ in average. This implies a vertical widening of 12.6 km
in just $\sim18$ yr, of an initially flat distribution of particles.
\item The Pallene ring may be formed of particles of small eccentricity, 
below $\sim$0.006.
Despite the larger size of Pallene compared to Aegaeon, Methone
or Anthe, the small moon is unable to efficiently clean up its orbit, 
due to its anyway small size and large $i$. The average final 
inclination of ring particles is $0.001023^\circ$,
which leads to a vertical width of $\sim$3.8 km.
\end{itemize}

All the regions explored in this work have some stable zones
where no particles are found in the real system. If ever some particles
existed in such regions, then some mechanism should be
responsible for their removal. Such mechanisms are most likely the
non-gravitational forces, such as solar radiation force and plasma
drag, as it is expected that particles close to the small moons were
originated when micrometeoroid IDPs hit
the surfaces of the small moons, leaving ejecta that is also 
micrometrical in size.\\

The influence of non-gravitational forces have been studied for
the regions of Aegaeon (for example in Madeira et al.,
in prep.), Methone and Anthe \citep{Sun17}. The work exploring
such shaping forces for the Pallene ring is under investigation.

\section*{Acknowledgements}

We acknowledge an anonymous referee for insightful comments that help
to improve the present work. We thank the financial support from FAPESP 
(Proc. N$\textsuperscript{o}$ 2016/01467-8 
and 2011/08171-3). SGW also thanks CNPq project number 309254/2012-4. 
We acknowledge the use of the 
Saturn Cluster of the Grupo de Din\^amica Orbital e Planetologia at UNESP, 
campus of Guaratinguet\'a.

%%%%%%%%%%%%%%%%%%%%%%%%%%%%%%%%%%%%%%%%%%%%%%%%%%



\bibliographystyle{mnras}
\bibliography{SatBibliography} 


%%%%%%%%%%%%%%%%%%%%%%%%%%%%%%%%%%%%%%%%%%%%%%%%%%

%%%%%%%%%%%%%%%%% APPENDICES %%%%%%%%%%%%%%%%%%%%%

%\appendix


%%%%%%%%%%%%%%%%%%%%%%%%%%%%%%%%%%%%%%%%%%%%%%%%%%

% Don't change these lines
\bsp	% typesetting comment
\label{lastpage}
\end{document}